\documentclass{aa}
\usepackage[varg]{txfonts}
\usepackage{hyperref}
\usepackage{lipsum}
\usepackage{amsmath}
\usepackage{txfonts}
\usepackage[utf8]{inputenc}
\usepackage[utf8]{luainputenc}
\usepackage{babel}
\usepackage{graphicx}
\usepackage{amssymb}
\usepackage{lipsum}
\usepackage{float}

\usepackage{bm}

\usepackage{import}
\usepackage{siunitx}
\DeclareSIUnit \parsec {pc}
\DeclareSIUnit \megaparsec {Mpc}
\DeclareSIUnit \year {yr}
\DeclareSIUnit \erg {erg}
\usepackage[plain]{algorithm}
\usepackage{caption}
\usepackage[noend]{algpseudocode}

\usepackage{acronym}

\newcommand{\bb}[1]{\left(#1\right)}                                        
\newcommand{\bc}[1]{\left[#1\right]}                                        
\newcommand{\bs}[1]{\left\{#1\right\}}                                      
\newcommand{\abs}[1]{\left|#1\right|}                                       

\newcommand{\kB}[0]{k_{\mathrm{B}}}                                         

\newcommand{\aref}[1]{Appendix~\ref{sec:#1}}
\newcommand{\eref}[1]{Eq.~\ref{eq:#1}}
\newcommand{\equref}[1]{Equation~\ref{eq:#1}}
\newcommand{\fref}[1]{Fig.~\ref{fig:#1}}
\newcommand{\figref}[1]{Figure~\ref{fig:#1}}
\newcommand{\secref}[1]{Section~\ref{sec:#1}}
\newcommand{\sref}[1]{Section~\ref{sec:#1}}

\newcommand{\knusbar}[0]{\bar{k}_{\nu,\mathrm{s}}}
\newcommand{\knus}[0]{k_{\nu,\mathrm{s}}}
\newcommand{\knuabar}[0]{\bar{k}_{\nu,\mathrm{a}}}

\newcommand{\trec}[0]{t_{\mathrm{rec}}}
\newcommand{\direc}[0]{\bm{\Omega}}

\newcommand{\Rstroemgren}[0]{R_{\mathrm{St}}}

\newcommand{\Rr}[0]{R_{\mathrm{r}}}

\newcommand{\DeltaT}[0]{\Delta t_{\mathrm{max}}}
\newcommand{\outf}[0]{f}
\newcommand{\timestepFactor}[0]{x}
\newcommand{\protonMass}{m_{\mathrm{p}}}
\newcommand{\RomanNumeral}[1]{\uppercase\expandafter{\romannumeral #1\relax}}
\newcommand{\neutralHydrogen}{H \, {\RomanNumeral{1}}}
\newcommand{\ionizedHydrogen}{H \, {\RomanNumeral{2}}}
\newcommand{\neutralHydrogenDensity}{n_{\mathrm{\neutralHydrogen}}}
\newcommand{\ionizedHydrogenDensity}{n_{\mathrm{\ionizedHydrogen}}}
\newcommand{\electronDensity}{n_{\mathrm{e}}}
\newcommand{\hydrogenDensity}{n_{\mathrm{H}}}
\newcommand{\xhii}[0]{x_{\mathrm{\ionizedHydrogen}}}

\bibpunct{(}{)}{;}{a}{}{,} 

\begin{document}

\title{Subsweep: Extensions to the sweep method for radiative transfer}

\author{
Toni Peter
\inst{1}
\and
Joseph S.\ W.\ Lewis\inst{2}
\and
Ralf S.\ Klessen\inst{1},\inst{3}
\and
Simon C.\ O.\ Glover\inst{1}
\and
Guido Kanschat\inst{3}
}

\institute{Universit\"{a}t Heidelberg, Zentrum f\"{u}r Astronomie, Institut f\"{u}r Theoretische Astrophysik, Albert-Ueberle-Str.\ 2, 69120 Heidelberg, Germany
\and
Sorbonne Université, Institut D`Astrophysique De Paris, 98 bis Boulevard Arago, 75014 Paris
\and
Universit\"{a}t Heidelberg, Interdisziplin\"{a}res Zentrum f\"{u}r Wissenschaftliches Rechnen, Im Neuenheimer Feld 225, 69120 Heidelberg, Germany \\
}

\date{Received September 15, 1996; accepted March 16, 1997}

\abstract{
  We introduce the radiative transfer postprocessing code {\sc Subsweep}.
 The code is based on the method of transport sweeps, in which the exact solution to the scattering-less radiative transfer equation is computed in a single pass through the entire computational grid. The radiative transfer module is coupled to radiation chemistry, and chemical compositions as well as temperatures of the cells are evolved according to photon fluxes computed during radiative transfer. {\sc Subsweep} extends the method of transport sweeps by incorporating sub-timesteps in a hierarchy of partial sweeps of the grid. This alleviates the need for a low, global timestep and as a result {\sc Subsweep} is able to drastically reduce the amount of computation required for accurate integration of the coupled radiation chemistry equations. We succesfully apply the code to a number of physical tests such as the expansion of HII regions, the formation of shadows behind dense objects, and its behavior in the presence of periodic boundary conditions.
}

\keywords{
radiative transfer -- methods: numerical -- (ISM:) HII regions
}

\maketitle

\section{Introduction}
Reionization is the process in which the universe shifts from being fully neutral to being almost completely ionized everywhere.
This is an important part of the transition from the primordial, homogeneous universe to the present-day universe which is full of heterogeneous, complex structures \citep[see for example][]{wiseCosmicReionisation2019,loebReionizationUniverseFirst2001,zaroubiEpochReionization2013}. We want to understand reionization by performing a set of simulations in which radiative transfer and radiation chemistry are evolved in large cosmological simulations such as the Illustris-TNG simulations \citep{nelsonIllustrisTNGSimulationsPublic2021,naimanFirstResultsIllustrisTNG2018,springelFirstResultsIllustrisTNG2018,marinacciFirstResultsIllustrisTNG2018,nelsonFirstResultsIllustrisTNG2018,pillepichFirstResultsIllustrisTNG2018,nelsonFirstResultsTNG502019,pillepichFirstResultsTNG502019}. Ideally, one would like to model reionization ``on-the-fly'' in such simulations, that is at the same time as the rest of the galaxy formation physics, since radiative feedback from reionization is thought to affect galaxy formation \citep{shapiroPhotoevaporationCosmologicalMinihaloes2004}. Moreover, the intergalactic medium reacts to the passage of reionization fronts possibly resulting in the disruption of filaments \citep{daloisioHydrodynamicResponseIntergalactic2020}.

However, the computational cost for doing so in a large, high-resolution simulation can be extremely large \citep{Gnedin2022}. There is therefore considerable interest in finding efficient ways to model reionization in a post-processing step that can be applied to such simulations. While such an approach misses the recoupling of radiative feedback to hydrodynamics, it does allow us to compare the results of the simulation with and without radiative feedback without having to perform an additional, expensive hydrodynamics simulation.

Here, we introduce a new code, {\sc Subsweep}, which is designed to solve radiative transfer and radiation chemistry on large inputs, such as those from cosmological simulations, and which is therefore ideally suited for this application.

Radiative transfer is the physical theory that describes the propagation of radiation and its interactions with matter, such as absorption and scattering~\citep{mihalasFoundationsRadiationHydrodynamics1999}.
There are multiple reasons why radiative transfer is a particularly challenging numerical problem, beginning with the high dimensionality of the quantity of interest: specific intensity, which depends on three spatial, two directional, one temporal, and one frequency dimension leading to a total of seven dimensions.
In addition, the radiative transfer equation can be seen as an elliptical equation in optically thick media and as a hyperbolic equation in optically thin media, making it particularly difficult to select a single solution method that works well across the entire parameter space.
Furthermore, we are primarily interested in physical scenarios in which the properties of the medium in which the radiation is transported change due to the influence of the radiation. At the same time, the radiation transport itself is dependent on the properties of the underlying medium (such as emissivity and opacity), calling for a solution to the coupled equations of radiative transfer and radiation chemistry.

Moment-based methods, where one solves the moments of the radiative transfer equation with some approximate closure relation, as opposed to the full radiative transfer equation, are a leading class of methods.
This approximation can lead to drastically improved performance while decreasing the overall accuracy of the result.
All moment-based methods need to choose a closure relation, which is typically given in terms of an expression for the Eddington tensor.
One of the first examples of a moment-based method is the flux-limited diffusion approach~\citep{levermoreFluxlimitedDiffusionTheory1981, whitehouseSmoothedParticleHydrodynamics2004} in which one assumes that the intensity varies sufficiently slowly in time and space that the radiative transfer equation becomes an effective diffusion law for the photon fluxes, and introduces a flux limiter to ensure that the signal speed of the radiation field remains lower than the speed of light. Flux-limited diffusion has been successfully applied in astrophysics \citep[for example][]{krumholzRadiationHydrodynamicSimulationsCollapse2007,bossFluxlimitedDiffusionApproximation2008}, but its main drawback is its diffusive nature which results in a lack of proper shadow formation \citep[see for example][]{hayesFluxlimitedDiffusionParallel2003}.

Another example of a moment-based method is the optically thin variable Eddington tensor method in which the Eddington tensor is computed under the assumption that all lines of sight to the sources in the simulation are optically thin \citep{gnedinMultidimensionalCosmologicalRadiative2001}. While efficient, this algorithm is applicable only to a comparatively narrow range of problems.

The M1 method is another moment-based method based on the M1 closure relation~\citep{aubertRadiativeTransferScheme2008,kannanArepoRTRadiationHydrodynamics2019}. While it is comparatively fast, it suffers from numerical problems inherent to moment-based methods, such as the two-beam instability~\citep{rosdahlRamsesrtRadiationHydrodynamics2013}.

Finally, a closure relation can also be derived by explicitly computing the Eddington tensor using a ray-tracing approach \citep[for example][]{hayesFluxlimitedDiffusionParallel2003,Finlator2018}. Such methods are extremely accurate but are also typically far too slow to apply to high resolution simulations (e.g.\ the \citealt{Finlator2018} calculation solved the radiative transfer problem on a coarse $64^3$ grid, a factor of $\sim 20000$ smaller than the effective resolution of even the smallest TNG simulation).

Another class of methods is given by Monte Carlo approaches, in which radiation is represented by individual photon packets~\citep{oxleySmoothedParticleHydrodynamics2003,dullemondRADMC3DMultipurposeRadiative2012}. For each source of radiation, packets are created by sampling over the direction of the emitted radiation (and its frequency, in multi-frequency versions of this approach) according to an appropriate probability distribution. The packets are then propagated through the gas and allowed to interact with it. These interactions are also treated probabilistically: the optical depth of a given fluid element can be used to derive the probability that a photon packet that encounters it will be absorbed. Scattering can be treated in a similar fashion.
A primary advantage of these methods is that their accuracy is determined by the number of emitted photon packets, making it easily tunable. Inherent to the nature of the method is statistical noise, with a signal-to-noise ratio that scales as $\mathrm{SNR} \propto \sqrt{n}$, where $n$ is the number of photon packets. However, an important disadvantage of this type of method is that it is difficult to parallelize when applied to very large simulations. For small simulations, the optimal parallelization strategy is typically to duplicate the computational domain on each processor and to split up the work by sharing the total number of photon packages evenly between the processors. Provided that the problem is small enough to all this, this approach provides close to linear scaling up to 100s or 1000s of cores \citep[see for example][]{hyperion}. However, once the problem becomes too big to allow the whole computational domain to be held in the memory available to a single processor, this simple strategy is no longer possible, and efficient parallelization of the algorithm becomes much harder. Moreover, the cost of this approach necessarily scales with the number of sources in the simulation, which can quickly become a problem for cosmological simulations containing millions of galaxies. 

In \citet{peterSweepMethodRadiative2023}, we introduced the sweep method and its implementation in {\sc Arepo}, {\sc Sweep}, which is a very efficient way of computing the exact solution to the scattering-less radiative transfer equation in parallel based on the concept of transport sweeps~\citep{kochSolutionFirstorderForm1991,zeyaoParallelFluxSweep2004}.
Transport sweeps are a subclass of discrete ordinate methods, in which the radiative transfer equation is solved simply by discretizing it in all of the available variables (time, space, frequency, and angle).
During a sweep, scattering is assumed to be negligible, such that the radiative transfer equations for different angles decouple (note that scattering can still be computed correctly by reintroducing it as a source for subsequent sweep iterations).
In order to solve the resulting equations efficiently, the algorithm computes an ad-hoc topological sorting of the grid with respect to the direction of the sweep, resulting in a method that computes the exact solution to the equation in a single pass through the grid. We have found the resulting method to be very performant and accurate at the same time.

Our previous implementation of the sweep method within the cosmological simulation code {\sc Arepo} \citep{springelPurSiMuove2010}, {\sc Sweep} \citep{peterSweepMethodRadiative2023}, worked well on medium sized inputs, but a major problem was the fact that it could only perform global operations in which radiative transfer is performed on the entire box, before a global chemistry update is performed. This limitation makes large runs prohibitively expensive since the need for a small timestep in one of the cells of the entire simulation will imply a small global timestep everywhere.

In this paper, we introduce the standalone radiative transfer postprocessing code {\sc Subsweep}, which deals with this problem by introducing a substepping procedure, similar to the sub-timestepping in modern hydrodynamical codes.
This method works by assigning grid cells individual timesteps, which are chosen from a power-of-two hierarchy and adapted to the local, physical timescales of processes relevant to radiative transfer.
Transport sweeps are then performed according to this timestep assignment in a physically consistent way, resulting in a computation in which cells with very low desired timesteps can be evolved accurately without sacrificing performance by strictly adhering to a global, low timestep.

We also find that this new method drastically alleviates the computational cost of incorporating periodic boundary conditions, one of the main challenges for the initial version of the sweep algorithm.
Previously, periodic boundary conditions were implemented by source iteration: photon fluxes leaving the simulation box are introduced as a source term for a subsequent sweep, until convergence is reached.
In the new substepping approach, we use the concept of warmstarting: periodic source terms from the previous iteration are used as a guess for the new timestep. By applying this concept also to the sub-timestep sweeps, we find that sufficient accuracy for our applications is reached without performing any additional source iterations.

This paper is structured as follows.
In \secref{methods}, we discuss the implementation details of {\sc Subsweep}, with a particular focus on the spatial domain decomposition (\secref{domainDecompSubsweep}) and the construction of the Voronoi grid (\secref{voronoiGridConstruction}). We also focus on radiative transfer in general and the sweep method in particular (\secref{sweep}) before we introduce the substepping approach (\secref{substepping}). \secref{methods} ends with the details of our radiation chemistry solver (\secref{chemistry}).
Afterwards, we perform a number of tests of our code (\secref{tests}), showing the physical accuracy of the results in an R-Type expansion test in the normal case (\secref{rtype}) and the case of the expansion happening across a periodic boundary (\secref{periodic}). We also perform a test to study the shadowing behavior of the code (\secref{shadowing}).
We assess the performance of the substepping method by performing a one-dimensional R-Type expansion for a large range of parameters (\secref{oneDExpansion}) and perform a brief series of tests for the radiation chemistry solver (\secref{chemistryTests}).
Finally, we conclude this paper and discuss future extensions of the code as well as possible applications in \secref{conclusion}.

\section{Methods}\label{sec:methods}
\subsection{General structure of the code}
In this paper we discuss the {\sc Subsweep} simulation code\footnote{Source code publicly available at \url{https://github.com/tehforsch/subsweep}} which is a standalone code for post-processing large cosmological simulations.
Currently, the code works with outputs of {\sc Arepo} simulations, but extensions for output formats of other simulation codes are possible.
The code requires data specifying the coordinates, temperatures and chemical compositions of the cells.
Source terms can either be explicitly specified by the user or will be computed from a set of source cells (such as star particles in the case of {\sc Arepo}) which also needs to be present in the inputs.
It will then distribute the data onto the desired number of cores (which we briefly discuss in \sref{domainDecompSubsweep}), construct a the Voronoi grid (discussed in \sref{voronoiGridConstruction}) and solve the radiative transfer equation coupled to radiation chemistry and write out the intermediate and final results of the computation.
Documentation for the usage of the code is available alongside the source code.
The code is written with a particular focus on the postprocessing of high-redshift cosmological simulations and reionization, but extensions incorporating present-day chemistry into the code are straightforward.

\subsection{Domain decomposition\label{sec:domainDecompSubsweep}}
In order to run our code in parallel, we have to distribute the available data over multiple cores.
We choose to use a simplified version of a standard Peano-Hilbert space-filling curve approach to spatial domain decompositioning which we will briefly describe in the following.

The problem that the domain decomposition tries to solve is to distribute the particles onto the $n$ cores $1 \dots n$ in such a way that the total runtime of the program is minimized.
Since this is a very difficult optimization problem to solve in general, we make it more concrete by defining two primary goals of the domain decomposition.
The first goal is the minimization of the load imbalance $\frac{\text{max} \{ L_{i} \} - \text{min} \{ L_{i} \}}{\text{max} \{ L_{i} \}}$ where the load $L_i$ on core $i$ can be defined in a variety of ways, which we will discuss later.

The second goal is to keep the total time spent communicating as low as possible.
This requirement is almost equivalent to minimizing the surface area of the intersection between the domains, because shared interfaces are where communication needs to take place in order to solve them.

A third priority that is specific to transport sweep algorithms is that even if goals 1 and 2 are fulfilled optimally, the resulting sweep can still be slow if the cells are arranged in such a way that not all cores can work simultaneously due to the task dependencies that need to be fulfilled (for more discussion, see \citealt{peterSweepMethodRadiative2023}).

For structured grids, a domain decomposition that optimizes the parallel performance of the transport sweep is given by the Koch-Baker-Alcouffe algorithm~\citep{bakerAlgorithmMassivelyParallel1998,kochSolutionFirstorderForm1991}.
For unstructured grids, optimizing the performance by the domain decomposition is difficult in general, which has been discussed in detail \citep{vermaakMassivelyParallelTransport2020, adamsProvablyOptimalParallel2019}.

Here, we will briefly discuss our implementation of a well-known approach based on space-filling curves that can solve requirements 1 and 2 simultaneously.
For now, we find that even though we do not optimize explicitly for the third goal, that is the sweep scheduling, the resulting performance is good enough for our purposes.

In our case, a space-filling curve is given by a mapping $f$ and its inverse $f^{-1}$ between a one-dimensional interval and all the possible floating point positions in the three-dimensional simulation box with side length $L$,
\begin{align}
  f&: [c_{\text{min}}, c_{\text{max}}] \rightarrow [0, L]^3, \\
  f^{-1}&: [0, L]^3 \rightarrow [c_{\text{min}}, c_{\text{max}}],
\end{align}
where $c_{\rm min}$ and $c_{\rm max}$ are the minimum and maximum values of the domain of the space filling curve respectively. We call $f(\bm{r})$ the key of a particle at position $\bm{r}$.
The basic idea of a domain decomposition using such a space filling curve is to move the three-dimensional optimization problem of distributing a set of points $\{p_j \in [0, L]^3\}$ onto $n$ cores $1 \dots n$ to a more tractable, one-dimensional problem.
In our case, this one-dimensional problem is the problem of finding cut-offs $s_i$ for $i = 1 \dots n-1$ so that the load imbalance is minimized if each core $i$ gets assigned the points $\{p_j \; \vert \; s_{i-1} < f^{-1}(p_j) < s_i \}$, where we take $s_0 = c_{\text{min}}$ and $s_n = c_{\text{max}}$.
If the space-filling curve is chosen such that it maps close-by points on the interval $[c_{\text{min}}, c_{\text{max}}]$ to close-by points in three dimensional space, the resulting distribution of points will form reasonably compact domains.
A common choice for such a curve is the Hilbert curve.

In order to execute the domain decomposition using the Hilbert curve, we require the load function $L_{i}(c_1, c_2)$ which computes the total load of the particles on core $i$ between the keys $c_1$ and $c_2$. Here, we assume that the load can be computed as a sum over the load for each particle.

\begin{algorithm}
  \caption{Cut-off search}\label{alg:cutoffSearch}
  \begin{algorithmic}[1]
    \Procedure{Find $s_i$}{}
    \State Initial guess: $s_{i} \gets s_{i-1}+\frac{c_{\text{max}} - s_{i-1}}{n-i}$
    \For{$d \gets 1, d_{\text{max}}$}
    \State For each rank $k$, compute $L_{k}(s_{i-1}, s_i)$.
    \State Compute $L \gets \sum_{k=1}^n L_{k}(s_{i-1}, s_i)$ via a global sum.
    \If{$L = L_{\text{local}}$}
    \Return $s_i$
    \ElsIf{$L < L_{\text{desired}}$}
    \State $s_i \gets \frac{s_i + c_{\text{max}}}{2}$
    \ElsIf{$L > L_{\text{local}}$}
    \State $s_i \gets \frac{s_i + s_{i-1}}{2}$
    \EndIf
    \EndFor
    \Return $s_i$
    \EndProcedure
  \end{algorithmic}
\end{algorithm}

In order to find the distributions of the cut-offs $s_i$ we proceed as follows:
\begin{enumerate}
    \item For each core, compute the keys for all local particles and sort them, so that computing the load function $L_{i}(c_1, c_2)$ becomes a cheap operation for any keys $c_1, c_2$.
    \item Compute the total load of the entire simulation $L_{\text{total}} = \sum_{i=1}^n L_{i}(c_{\text{min}}, c_{\text{max}})$ (via a global sum).
    \item Compute the desired load on each core as $L_{\mathrm{desired}} = \frac{L_{\text{total}}}{n}$.
    \item Using this, compute the cut-offs $s_i$, starting with $s_1$ using the parallel search described in Algorithm~\ref{alg:cutoffSearch}.
\end{enumerate}

\subsection{Construction of the Voronoi grid\label{sec:voronoiGridConstruction}}
In order to perform the sweep algorithm over a set of points, we need to construct a mesh, so that we can determine the connectivity of cells.
In order to avoid any additional numerical artifacts, we decided to use a similar mesh to the one that was used in the code which generated the outputs which we are trying to post-process using {\sc Subsweep}.
Since we are mostly interested in postprocessing simulation outputs of {\sc Arepo}, we choose to use a Voronoi grid, which is the mesh that {\sc Arepo} is based on.

There are many different algorithms for constructing Voronoi grids. For simplicity, the one here is based closely on the method described in \citet{springelPurSiMuove2010}. The method is based on incremental insertion~\citep{bowyerComputingDirichletTessellations1981,watsonComputingNdimensionalDelaunay1981}, extended to allow construction of the grid for a point set distributed onto multiple cores.

\subsubsection{Construction of the local Delaunay triangulation}
The Voronoi grid is constructed from its dual, the Delaunay triangulation. The serial incremental insertion algorithm for constructing the Delaunay triangulation proceeds as follows:
Given a set of $N$ mesh-generating points $\{p_i \; \vert \; 1 \leq i \leq N \}$, begin with an all-encompassing tetrahedron, that is, one that is large enough to contain all points $p_i$.
Now, for every point $p$, locate the tetrahedron in the triangulation which contains $p$.
How exactly this is done in a performant way is described in \secref{pointLocation}.
Using $p$, we split the tetrahedron containing $p$ into $4$ new tetrahedra.
After the split, the resulting triangulation is not necessarily Delaunay.
In order to restore the Delaunay property, we begin by putting each of the $4$ newly formed tetrahedra on a stack.
For each tetrahedron $t$ in the stack, we find the face $F$ which is opposite of $p$ in $t$.
We then find the tetrahedron $t'$ which is on the other side of $F$, and locate the point $p'$ which is opposite of $F$ within $t'$.
If $p'$ is contained in the circumcircle around $t$, then the face $F$ violates the Delaunay criterion and needs to be removed.
To do so we perform a flip orientation on the two tetrahedra $t$ and $t'$ which will result in a number of new tetrahedra, each of which will now have to be checked for the Delauny property, so we put them on the stack as well.
Once the stack is empty, the Delaunay property has been restored again and we can begin inserting the next point.

The flip operation between two tetrahedra $t$ and $t'$, their shared face $F$ and the two points $p$ and $p'$ opposite of $F$ in each of the tetrahedra respectively works as follows:
Compute the intersection point $q$ of the face $F$ with the line between $p$ and $p'$.
If $q$ lies inside $F$, we perform a 2-to-3-flip, in which the two tetrahedra are replaced by three.
If the intersection point lies outside one of the edges of $F$, we take into account the neighboring tetrahedron along that edge and perform the opposite operation - a 3-to-2 flip - in which the three tetrahedra are converted to two.
If the intersection point lies outside two edges, the flip can be skipped.
It can be shown \citep{edelsbrunnerIncrementalTopologicalFlipping1996} that flipping the remaining violating edges will restore the Delaunay property.
For more information on this procedure see \citet{springelPurSiMuove2010}.

\subsubsection{Point location\label{sec:pointLocation}}
While inserting a point $p$ into the triangulation, we need to locate the tetrahedron containing $p$. This is performed by the simple ``jump and walk" method.
The method works by using a priority queue $q$. We initialize $q$ as containing only the last tetrahedron that was inserted into the triangulation.
Now we iteratively take the highest priority tetrahedron $t$ out of the queue.
If $t$ contains $p$, return $t$.
If $t$ does not contain $p$, we add all the neighboring tetrahedra of $t$ to $q$ with their priority determined by their distance to $p$ (so that tetrahedra closest to $p$ are searched first).
The method performs best if the order of the points inserted into the triangulation is such that two points inserted after another are also at similar positions (which in turn makes the initial guess better). In order to achieve this, we begin the construction by sorting all points according to their Peano-Hilbert key.

\subsubsection{Parallel Delaunay construction\label{sec:delaunayParallel}}
In principle, we would like to construct the global Delaunay triangulation $T_{\text{global}}$ on all of the points in the entire simulation.
In practice, we are limited to those points that are available on each core.
All we can do is to construct a local triangulation $T_{\text{local}}$ over all of the local points.
The goal of the triangulation is to provide connection information and in order for it to be useful, this connection information has to be consistent with what the other cores see.
It is clear that in order to do so and preserve the Delaunay property we need to import points that lie on other cores, which we call halo points.
More precisely, we want to construct $T_{\text{local}}$ in such a way that it is consistent with $T_{\text{global}}$, in the sense that for every local point $p$, the set of tetrahedra $\{t \mid t \in T_{\text{local}}, p \in t\}$ is the same as $\{t \mid t \in T_{\text{global}}, p \in t\}$. Note that this requirement does not extend to halo points, allowing us to stop importing additional halo points once all local points are consistent in the above sense.

Here, we will describe the algorithm for the halo search.
The goal here is to import every necessary halo point in order to reach a consistent local triangulation, while importing as few as possible superfluous points in order to speed up the grid generation and keep memory overhead as low as possible.
The basic idea is that a tetrahedron $t$ is consistent with the global triangulation $T_{\text{global}}$ iff we have imported the set of all points $\{p \mid p \in C(t)\}$ from all other cores, where $C(t)$ is the circumcircle of the tetrahedron.
To do so, we begin by constructing the Delaunay triangulation of all local points $T_{\text{local}}^0$.
Initially, we flag every tetrahedron in the triangulation as ``undecided" and then iterate on the following process:

For every undecided tetrahedron $t$, we compute the circumcircle $C(t) = (c, r)$ with center $c$ and radius $r$. Given the circumcircle $C(t)$, we compute the search radius $r'$.
Search other cores for all points $p$ that are within $r'$ distance of $c$ which we have not imported locally yet.
If there is no such point anywhere (which means we have imported all points that could be contained in the circumcircle of the tetrahedron), flag the tetrahedron as ``decided".
Otherwise, we add all points $p$ to the list of newly imported points.
Now, we construct $T_{\text{local}}^{i+1}$ by inserting the set of all newly imported points into $T_{\text{local}}^i$.
We flag any newly created tetrahedron which contains a local point as undecided.

Here, the search radius $r'$ is computed as follows: If the radius of the circumcircle $r$ is smaller than the average expected size of a Voronoi tetrahedron $\bar{l}$, then $r' = r$.
Otherwise, we use $r' = \bar{l}$, unless we have previously performed a radius search for this tetrahedron before, in which case we use $r' = r'_{\text{previous}} * \alpha$ where $\alpha > 1$ is a free parameter.
This is done because in the first few iterations of the triangulation, very large tetrahedra tend to  form because we are not yet aware of the presence of very nearby points on other cores.
If we blindly performed a radius search with the radius of the circumcircle $r$, we might unnecessarily import a large number of points from other cores.
However, if the triangulation should contain this large tetrahedron, the exponentially increasing search radius will ensure that we perform a search with the proper radius within a reasonable number of iterations.

If periodic boundaries are desired, the same procedure described above, which imports halo points from other ranks can also import periodic haloes (points that represent a point shifted by a multiple of the box size along one or more axes) both from other cores and from the set of local points.
This can be done by modifying the radius search such that it takes periodic images of points into account.
Since constructing the distributed triangulation requires many radius searches, we need to perform the radius search quickly. To do so, we construct a standard Oct-tree on the set of all local points which reduces point search to a $\mathcal{O}(n \log{}n)$ operation.

\subsubsection{Degeneracies}
Another difficulty in creating Delaunay triangulation is how to deal with degenerate cases and those that are close to being degenerate. One solution to this problem is to perform all operations in arbitrary precision arithmetic. However, this will reduce the performance of the code drastically. In {\sc Subsweep} we take an approach similar to the one in~\citet{springelPurSiMuove2010} where we perform the critical checks (such as the checks that ask whether a point is contained in a tetrahedron or whether a tetrahedron is positively oriented) in floating point arithmetic first. If the result of the floating point operation is at risk of being qualitatively wrong due to numerical round-off errors, we perform it again in arbitrary precision arithmetic. In the current code, we do not deal with truly degenerate cases (for example, a point lying exactly on a face of a tetrahedron) because we find them to be extremely rare in practice, however, it is possible to extend the method to account for degeneracies. For more information on this procedure, we refer to~\citet{springelPurSiMuove2010}.

\subsection{The sweep algorithm\label{sec:sweep}}
We introduced the sweep algorithm in the context of astrophysics in \citet{peterSweepMethodRadiative2023}.
Here, we will briefly recap the basics of the algorithm in order to explain the required fundamentals for understanding the extensions we will introduce in later sections.

Given the specific radiative intensity $I_{\nu}(\bm{r}, t, \hat{\Omega})$, with frequency $\nu$, spatial position $\bm{r}$, time $t$, and solid angle $\hat{\Omega}$ given in units of $\SI{}{\watt\per\meter\squared\per\steradian\per\hertz}$, the general radiative transfer equation (RTE) reads \citep{rybickiRadiativeProcessesAstrophysics1985}
\begin{align}
\frac{1}{c}\frac{\partial}{\partial t}I_\nu + \hat{\bm{\Omega}} \cdot \bm{\nabla} I_\nu = j_\nu - (\knusbar+\knuabar) I_\nu + \frac{1}{4\pi} \int_{S} \knus(\bm{\Omega}') I_\nu \bm{\mathrm{d}\Omega'}. 
\label{eq:radiativeTransferGeneral2}
\end{align}

In the case of {\sc Subsweep}, we assume that scattering terms are negligible (see \citet{peterSweepMethodRadiative2023} for more details) and make the infinite speed of light assumption, so that we obtain
\begin{align}
\hat{\bm{\Omega}} \cdot \bm{\nabla} I_\nu = j_\nu - \knuabar I_\nu. 
\label{eq:radiativeTransfer2}
\end{align}
We note here that the infinite speed of light assumption is indeed quite a strong assumption in cosmological contexts, where light-crossing times of large intergalactic structures can quickly become relevant, so that assuming an infinite speed of light can cause ionization fronts to be affected by a change in the source population more quickly than they would in reality. However, this only has a minor impact on the overall evolution of the system. For more discussion of the impact of an infinite speed of light in a cosmological context, see for example \citet{kahou2023}.

The sweep method is a discrete ordinate method, which means that it solves the RTE by discretizing it in every variable, that is in space, time, angle and frequency.
\begin{figure}
    \centering
    \includegraphics[width=\columnwidth]{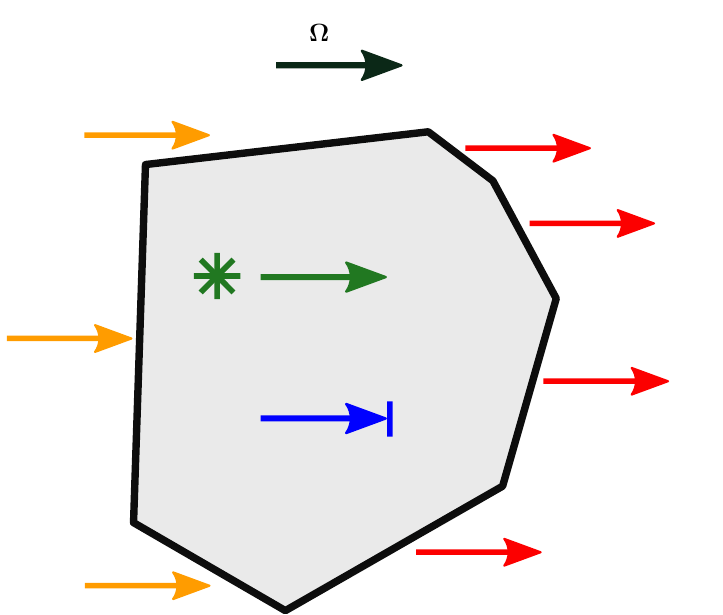}
    \caption{Illustration of the radiative processes described by \eref{radiativeTransfer2} for a single grid cell. The processes are: Incoming (orange) and outgoing (red) radiation, sources (green), absorption (blue)\label{fig:radiativeProcesses2}}
\end{figure}

\equref{radiativeTransfer2} can be intuitively understood using the illustration in \fref{radiativeProcesses2}, which shows the processes affecting a small volume of space.
Sources of radiation in this volume are through incoming radiation from cells to the left (orange arrows), the source term $j$ directly (green arrow). Radiation from the cell is either absorbed (blue arrow) or leaves the cell towards the right (red arrows).
The neighboring cells therefore fall into two categories. Cells upwind of the cell along $\direc$ (orange arrows) need to have their solution computed before this cell, since we require the incoming fluxes from those cells to solve the local problem. Cells downwind of the cell require the outgoing fluxes of the local solution in order to be solved.

The crucial idea of the sweep method is that it finds a topological sorting of the partial order induced by the upwind-downwind relation, such that the exact solution to the (scattering-less) RTE can be obtained in $n_{\mathrm{dir}}$ passes through the grid, where $n_{\mathrm{dir}}$ is the number of bins into which we choose discretize the angular directions.
It is crucial that the upwind-downwind relation is transitive, so that it is a partial order (which is equivalent to there being no cycles in the directed graph induced by the order), because otherwise a topological sorting of the cells does not exist. 
We have shown in \citet{peterSweepMethodRadiative2023} that this is always true for such an ordering induced by a Voronoi grid, which is the only type of grid that we are going to work with in this paper.
 It should be noted that the ordering is trivially acyclic for Euclidean grids or grids generated by adaptive mesh refinement, so that the sweep algorithm can also be used in other types of grid that are widely used in astrophysics.

\begin{algorithm}
\caption{Single-core sweep}\label{alg:serialSweep}
\begin{algorithmic}[1]
\State initialize task queue $q \gets \bs{}$
\For{all $\direc$ and all cells $c$ in grid}
\State count number of required upwind fluxes $n(c, \direc) \gets u(t)$
\State\textbf{if} {$n(c, \direc) = 0$} \textbf{then} add task $(c, \direc)$ to $q$
\EndFor
\While{$q$ not empty}
\State get first task $t = (c, \direc)$ from $q$
\State solve $t$ using upwind fluxes \label{step:solver2}
\For{downwind neighbor $c_{\mathrm{d}}$ in $d(t)$}
\State reduce missing upwind flux count $n(c_{\mathrm{d}}, \direc)$ by 1.
\State\textbf{if} {$n(c_{\mathrm{d}}, \direc)$ = 0} \textbf{then} add task $(c_{\mathrm{d}}, \direc)$ to $q$.
\EndFor
\EndWhile
\end{algorithmic}
\end{algorithm}

The single-core sweep algorithm is described in Algorithm~\ref{alg:serialSweep}. In order to find the topological sorting, the sweep algorithm starts the computation by computing an upwind count for each direction and each cell which is simply the number of cells upwind of the cell in the given direction.
The idea is to keep track of the set of all (cell, direction) pairs which can currently be solved, which are those whose upwind neighbors have already been solved.
Whenever we solve a task, we reduce the upwind count of all its downwind dependencies by 1. If the upwind count of this dependency is now zero, we put this dependency into the task queue.
Once the queue is empty, we have solved all tasks and have obtained the solution to the RTE.
The grid being acyclic guarantees that this algorithm always terminates.

\begin{algorithm}
\caption{Parallel sweep}\label{alg:parallelSweep2}
\begin{algorithmic}[1]
\State initialize task queue $q \gets \bs{}$
\State initialize send queues for each processor $i$ holding downwind neighbors of any of the cells in the domain of the current processor: $s_i \gets \bs{}$
\For{all $\direc$ and all cells $c$ in grid}
    \State count number of required upwind fluxes $n(c, \direc) \gets u(t)$
    \State \textbf{if} {$n(c, \direc) = 0$} \textbf{then} add task $(c, \direc)$ to $q$
\EndFor
\While{any cell unsolved or any $s_i$ not empty}
    \For{each incoming message (flux $f$ along $\direc$ into cell $c$)}
        \State reduce missing upwind flux count $n(c, \direc)$ by 1.
        \State \textbf{if} {$n(c, \direc)$ = 0} \textbf{then} add task $(c, \direc)$ to $q$.
    \EndFor
    \State $n_{\mathrm{solved}} = 0$
    \While{$q$ nonempty \textbf{and} $n_{\mathrm{solved}} < N_{\mathrm{max}}$}
        \State get first task $t = (c, \direc)$ from $q$
        \State solve $t$ using upwind fluxes
        \State $n_{\mathrm{solved}}$ += 1
        \For{downwind neighbor $c_{\mathrm{d}}$ in $d(t)$}
            \If{$c_{\mathrm{d}}$ is remote cell on processor $i$}
                \State add flux to send queue $s_i$ \label{step:sendQueue2}
            \Else 
                \State reduce missing upwind flux count $n(c_{\mathrm{d}}, \direc)$ by 1.
                \State \textbf{if} {$n(c_{\mathrm{d}}, \direc)$ = 0} \textbf{then} add task $(c_{\mathrm{d}}, \direc)$ to $q$.
            \EndIf
        \EndFor
    \EndWhile
    \State send all messages in $s_i$
\EndWhile
\end{algorithmic}
\end{algorithm}

In order to perform the algorithm described in Algorithm~\ref{alg:serialSweep} in parallel on many cores with a spatial domain decomposition, a number of modifications need to be made to the algorithm.
The basic idea of the algorithm does not change in the parallelized version of the code.
The main difference is that we need to take task dependencies between different cores into account.
Previously, having solved a task meant that we could simply reduce the upwind count of all its downwind dependencies by one.
Now, the downwind dependencies of a task might be on a different core.
In this case, we send a message to that core consisting of the outgoing fluxes of the local cell, the ID of the downwind cell, and the direction of the task.
As soon as the other core receives that message, it will reduce the upwind count for the corresponding cell by one (and add it to the solve queue if the upwind count is 0 at this point).
In order to improve performance, messages are not sent immediately, since sending lots of small messages tends to increase communication overhead and reduce performance as a result.
The opposite strategy of sending messages only after all tasks that are solvable locally have been solved also comes with performance drawbacks, since it can cause long waiting times on neighboring cores who cannot perform any work before receiving new incoming fluxes.
In practice, we use an intermediate approach, where we solve at most $N_{\mathrm{max}}$ local tasks before new messages are sent and received.
Here, $N_{\mathrm{max}}$ is a free parameter and the two extremes are recovered for $N_{\mathrm{max}} = 1$ and $N_{\mathrm{max}} = \infty$, respectively.
In order for this parallel algorithm to terminate, it is crucial that all cores agree on the connections between their local cells.
If this is not the case, cores can end up waiting for incoming messages that will never be sent, causing infinite deadlocks.
The property that all cores agree on the interfaces between their boundary cells is ensured by the particular way in which grid construction is performed by the algorithm described in \secref{delaunayParallel}.

\subsection{Substepping\label{sec:substepping}}
\subsubsection{Motivation}
The main problem with the sweep algorithm above is that the entire grid operates on the same timestep.
This is not a problem for the RTE alone, since, given our assumption of infinite speed of light, it is independent of time and therefore reaches a steady-state solution immediately. However, we are interested in solutions of the RTE coupled to the radiation chemistry equations, which are manifestly time-dependent.

In practice, a large fraction of the cells in the simulation are either fully ionized or fully neutral and have settled into an equilibrium where their chemistry update could be performed on long timesteps. However, cells along ionization fronts require comparatively short timesteps in order to accurately integrate both the RTE and the chemical rate equations.
With the previous sweep algorithm, the only option was to use a low value for the global timestep, which in turn means that solving the system for the desired amount of time requires a higher number of sweeps.

\begin{figure}
    \centering
    \includegraphics[width=\columnwidth]{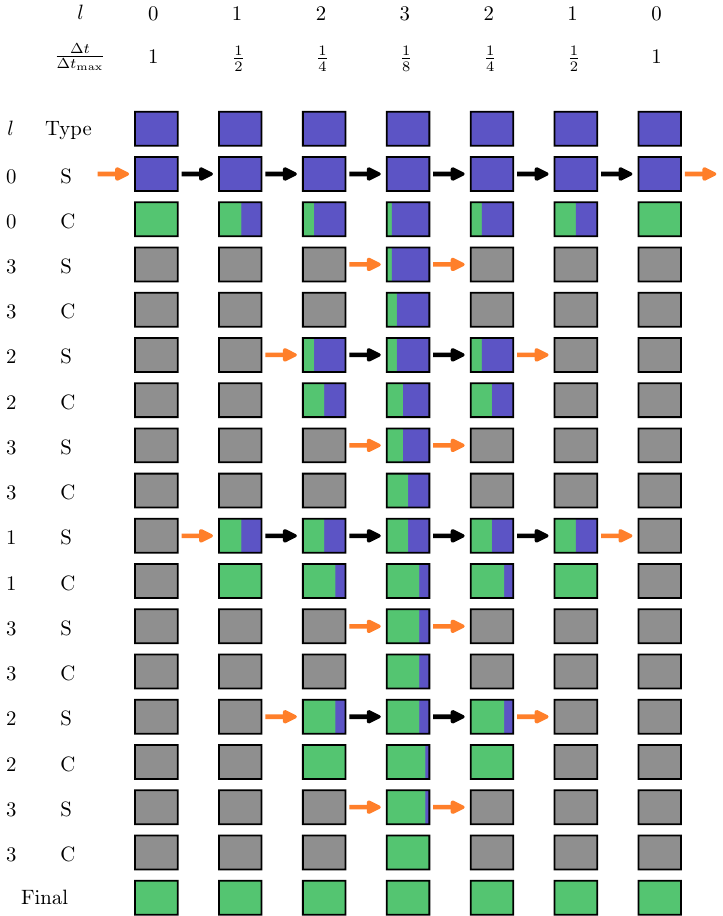}
    \caption{\label{fig:substepping1d}Illustration of the sweep substepping procedure. The rectangles represent cells, with the color of the cell indicating how far the cell has been integrated. A fully blue cell has not been integrated at all, while a green cell has been integrated for a total of $\DeltaT$. Arrows represent fluxes going into the cell which are computed during the sweep steps. Black arrows denote normal fluxes, while orange arrows represent boundary terms for the sweep. Each row represents either a sweep (denoted by S) or a chemistry update (denoted by C) at the corresponding level $l$. The last row represents the final state with each cell having been fully integrated.}
\end{figure}

In the following, we will introduce a modification to the sweep algorithm which allows cells to perform sub-timesteps, if required. Effectively, this solves the global timestep problem by letting cells choose their desired timestep. In the following, we will explain how this algorithm works in practice. An illustration of the algorithm is shown in \figref{substepping1d}.

\subsubsection{Timestep levels}
In order to do so, we introduce $n$ different timestep levels. Each cell $c$ is assigned to a timestep level $l(c)$. During a full timestep $\DeltaT$, a cell at timestep level $0$ receives one update for a full timestep $\DeltaT$. Cells at level $l$ receive $2^l$ updates with a timestep of $2^{-l} \DeltaT$.

At the end of each full timestep, each cell re-computes its desired timestep. To do so, we begin by computing the three timescales
\begin{enumerate}
    \item $t_{T}$, the timescale on which the temperature $T$ changes,
    \item $t_{x_{\mathrm{HII}}}$, the timescale on which the ionized hydrogen fraction $\xhii$ changes,
    \item $t_{F}$, the timescale on which the photon flux $F$ changes.
\end{enumerate}
Each individual timescale $t_q$ for the corresponding quantity $q$ (where $q$ is one of $T, \xhii$, and $F$) is computed as
\begin{align}
    t_q = \DeltaT 2^{-l} \left \vert \frac{q(t)}{q \left(t \right ) - q \left ( t - \DeltaT 2^{-l} \right ) } \right \vert
\end{align}
where $q \left ( t - \DeltaT 2^{-l} \right )$ and $q(t)$ refer to the values of $q$ in the previous partial sweep and the current one respectively and $l$ is the current timestep level of the cell.
Using these individual timescales, we define the minimum $t_{\mathrm{min}}$ as
\begin{align}
  \label{eq:changeTimescale}
  t_{\mathrm{min}} =\mathrm{min}\bc{t_{x_{\mathrm{HII}}}, t_{T}, t_{F}}.
\end{align}
Finally, the timescale $t_{\mathrm{min}}$ is used to compute the desired timestep $\Delta t_{\mathrm{desired}}$ for the cell as
\begin{align}
  \label{eq:timestepFactor}
  \Delta t_{\mathrm{desired}} = t_{\mathrm{min}} \timestepFactor,
\end{align}
where $\timestepFactor \in (0, 1]$ is a dimensionless free parameter which controls the accuracy of the integration.
Given $t_{\mathrm{desired}}$, we determine the timestep level $l'$ of the cell for the next full timestep as
\begin{align}
  \label{eq:timestepLevel}
    l' = \left\lceil \log_2 \frac{\DeltaT}{\Delta t_{\mathrm{desired}}} \right\rceil
\end{align}
for the entire next full timestep. In order to keep a fixed number of levels $n$, values of $l' > n-1$ are reduced to $n-1$ and values of $l' < 0$ are increased to zero.
Modifications of this method where cells can change their timestep level in the middle of a full timestep are possible, but for reasons of simplicity, we have not implemented them at this point.

\subsubsection{The algorithm}
Given the distribution of the cells onto the $n$ timestep levels $0 \dots n-1$, we introduce the following terminology:
A ``partial sweep'' at level $l$, or $l$-sweep is a sweep of all the cells which are at level $l$ or higher.
During a $l$-sweep, we call cells "active" if their timestep level $l'$ fulfills $l' \geq l$, that is if they are involved in the $l$-sweep.
A ``full sweep'' is the procedure by which the system is integrated for a full timestep $\DeltaT$ and consists of $1$ $0$-sweep, $1$ $1$-sweeps, $2$ $2$-sweeps, $4$ $3$-sweeps, \dots, and $2^{n-2}$ $\bb{n-1}$-sweeps. The order in which they are performed is illustrated in \fref{substepping1d}.

Since only the cells at levels larger or equal than $l$ participate in an $l$-sweep, we need to decide how to treat the incoming fluxes from cells which are at levels below $l$.
In this method, the fluxes of all cells at timestep levels $0 \dots l-1$ are kept constant and effectively treated as boundary conditions for the cells at higher levels.
During the $l$-sweep, we compute a flux correction for each cell that has at least one active upwind neighbor, since these are the only cells for which the fluxes change during this partial sweep. For these cells, the new outgoing flux $F_{\text{out}}'$ is computed as
\begin{align}
  \label{eq:fluxCorrection}
F_{\text{out}}' = F_{\text{out}} + \outf(F_{\text{in}}', c) - \outf(F_{\text{in}}, c),
\end{align}
where $F_{\text{out}}$ is the value of the outgoing flux before the partial sweep, $F_{\text{in}}$ is the incoming flux term before the partial sweep, $F_{\text{in}}'$ is the incoming flux in this partial sweep and $\outf$ is the function that computes the outgoing fluxes given the incoming fluxes and crucially depends on the chemical composition which might change during a chemistry update.
This function depends on the implementation of the chemistry, and its exact form in our hydrogen-only chemistry implementation will be discussed in \secref{chemistry}.
Outgoing fluxes of a cell are either used directly as input into other local cells, or communicated to other cores, as in the original sweep algorithm without substepping.
Flux corrections are applied to cells whether or not the target cell itself is active.

Once the $l$-sweep is finished, all cells $c$ have their chemistry updated by $ 2^{-l(c)}\DeltaT$.
This means that at this moment, cells on higher levels (and therefore shorter timesteps) have experienced ``less" time, than those on lower levels. This will be corrected by performing additional partial sweeps on the higher levels, so that at the end of a full sweep, each particle has been integrated for exactly $\DeltaT$. It should also be noted that consistency is guaranteed in the sense that the order in which the partial sweeps are performed guarantees that for any given partial sweep, all active cells have experienced the same amount of time.
After every full sweep, the cells are moved onto their new timestep level, according to their desired timestep (see \eref{timestepLevel}).
Crucially, after the timestep levels have been updated, each core communicates the new timestep level of each of its boundary cells to the neighboring cores. This is important because all of the cores have to agree on which cells are active at each level. If they do not agree on this, one of the cores will expect incoming fluxes over the interface shared by the two cells while the other will not send those fluxes, resulting in a deadlock of the partial sweep.

\subsection{Wind up}
At the beginning of the simulation, we do not know how to distribute the cells onto the timestep levels, since we have no prior data on the timescales at which their relevant quantities will change. If we had to guess the timestep level of any cell, the only reasonable choice we can make in order to not violate the timestep criterion of any single cell, is to place all cells in the highest level (the lowest timestep). However, performing a full sweep in this setup would require a total of $2^{n+1} - 1$ sweeps of the entire grid, an extremely expensive operation.
In order to avoid this, we compute the timescales of each cell by placing each cell in level $n-1$ and beginning with a $(n-1)$-sweep, that is a sweep at the smallest allowed timestep ($2^{-(n-1)} \DeltaT$). We then allow each cell to move one level down, if its desired timestep is large enough, and perform a $n-2$ sweep, and so on. In the end, we have performed $n$ partial sweeps and have simulated a total time of $\sum_{i=0}^n 2^{-i} \DeltaT = \bb{1 - 2^{-(n-1)}} \DeltaT$. In order to align the time intervals with multiples of $\DeltaT$, we perform one more partial sweep of all cells with timestep $2^{-(n-1)} \DeltaT$ which will bring the total simulated time to $\DeltaT$. From now on, every timestep will be performed by a full sweep, which totals $\DeltaT$.

\subsection{Periodic boundary conditions\label{sec:subsweep-pbc}}
Periodic boundary conditions are important in order to study cosmological volumes of space self-consistently, by allowing effects from the matter outside of the simulation box to be approximately modeled by the contents of the simulation box itself.
In the case of radiative transfer, this means re-introducing photons that escape the box on one side to the mirrored position on the opposite side.

In \sref{delaunayParallel}, we discussed how periodic boundaries are taken into consideration during mesh construction.
This means that each cell at the boundary of the box knows the location of its periodic neighbors.
There is no obvious, self-consistent way of re-introducing outgoing photons within a single sweep.
However, we can make use of the source iteration algorithm and treat fluxes that leave the boundaries of the simulation box as source terms for the next iteration of the algorithm.
Each iteration then approximates the true periodic source terms until convergence is reached.
However, applying this approach to a full sweep has the obvious drawback that every iteration takes exactly as long as the original sweep.
Since a full sweep over the grid is an expensive operation, repeating it a number of times in order to reach an acceptable level of convergence can quickly become infeasible.

In \citet{peterSweepMethodRadiative2023} we discussed the concept of warmstarting, where the resulting fluxes from previous sweeps are re-introduced in the next iteration in order to speed up convergence.
Moreover, warmstarting integrates extremely well with the sub-timestepping approach introduced in {\sc Subsweep}.
To do so, we use the outgoing periodic fluxes of every partial sweep as incoming fluxes into the corresponding cells for the next partial sweep.
This has a number of benefits.
Primarily, it changes the algorithm so that it does not require a global cost (re-running the full sweep) in order to fix an often local problem (convergence of the periodic fluxes in the cells with the most activity).
Instead, the algorithm naturally adapts itself to the local requirements and decreases the timestep in cells with particularly bad convergence behavior with respect to periodic boundary conditions.
It should be noted that this happens without requiring any timestep criterion specific to periodic boundary conditions: cells that have not converged to their true periodic fluxes will automatically reduce the timestep, since that derives (among other things) from the rate of change in the flux terms, as shown in \eref{changeTimescale}.
This combination of warmstarting and substepping has proven so effective that we have chosen not to implement any global iteration on levels of full sweeps in {\sc Subsweep}.

\subsection{Rotations}
As discussed for the original {\sc Sweep} implementation in \citet{peterSweepMethodRadiative2023}, we perform rotations of the directional bins between transport sweeps, similar to the approach described in \citet{krumholz2007}, in order to smooth out the effect that the discretization of the directional bins has on the result.
The preferential directions introduced by this discretization can easily lead to very apparent star-shaped artifacts in the hydrogen ionization fraction around strong sources.

In {\sc Subsweep}, we keep this approach to smoothing out preferential directions.
Here, remapping the flux corrections from one timestep to the next becomes important.
As in the original implementation, the directions $\bm{\Omega}_i$ are rotated to new directions $\bm{\Omega}_i = \bm{R}(\theta, \phi) \cdot \bm{\Omega}_i'$ where $\bm{R}(\theta, \phi)$ is the rotation matrix for the spherical coordinate-angles  $\theta$ and $\phi$.
The angles are chosen from a uniform distribution of $\theta \in [0, \pi]$, $\phi \in [0, 2 \pi]$.
Remapping of the fluxes onto the new angular bins is then done via $F(\bm{r}, \bm{\Omega}_i') = \sum_{j=1}^{N_{\mathrm{dir}}} \frac{\Delta S_{i j}}{\Delta S_i} F(\bm{r}, \bm{\Omega}')$ where $N_{\mathrm{dir}}$ is the number of directional bins, the interpolation coefficients $\Delta S_{i j}$ are given by the solid angle that $\bm{\Omega}_i$ and $\bm{\Omega}_j$ share, and $\Delta S_i$ is the solid angle corresponding to any direction $\Omega_i$.

These rotations are performed only after every full sweep and not after partial sweeps.
It is possible in principle to rotate the bins also after every partial sweep, but doing so can have a very strong, discontinuous effect on the convergence timescale of some cells.
In order to safely incorporate sub-timestep rotations into the substepping approach, we think it is necessary to introduce the ability for cells to change their desired timestep during partial sweeps, not only during full sweeps.
Therefore, we have chosen not to introduce this additional complexity to the algorithm.

The drawback of this choice is that if the full-sweep timestep $\Delta t_{\rm max}$ is chosen to be large compared to the timescales at which ionization fronts move a large amount of cells (which is desirable in order to fully take advantage of the substepping approach), artifacts due to preferential directions can be visible.
In order to avoid these artifacts, the full-sweep timestep has to be decreased, increasing computation time.

\subsection{Radiation chemistry\label{sec:chemistry}}
The implementation of the radiation chemistry in our code follows \citet{rosdahlRamsesrtRadiationHydrodynamics2013}.
In its current form, the code only treats the ionization, heating, and cooling of hydrogen in a primordial gas. We assume zero helium in our code.
However, extensions to incorporate helium or more complex chemical networks are possible and intended in the structure of the code.
This includes adding more frequency bins for the radiative transfer.
In the current form, we use one frequency bin which incorporates all photons with frequencies $\nu \geq f_{\mathrm{ion}}$, where $f_{\mathrm{ion}} = \frac{\SI{13.6}{eV}}{h}$.

The chemical state of a cell $c$ is described by the state vector: $U = (T, \xhii)$ alongside its (constant) density $\rho$.
The first thing that the implementation of the chemistry needs to provide is the function $\outf(F, c)$ discussed in \secref{substepping}. This function computes the outgoing photon flux of a cell $c$ given the incoming flux $F$ which depends on the chemical state $U$ of the cell. For our hydrogen-only chemistry, this function is given by
\begin{align}
  \label{eq:fluxEstimate}
  \outf(F, c) = F e^{-\neutralHydrogenDensity \sigma d},
\end{align}
where $\neutralHydrogenDensity$ is the density of neutral hydrogen, $d = \sqrt[3]{\frac{3 V}{4 \pi}}$ is the approximate size of the cell ($V$ is the volume of the cell), and $\sigma$ is the photon-number weighted average cross section of ionizing photons, defined as
\begin{align}
    \sigma = \frac{\int_{f_{\mathrm{ion}}}^{\infty} \sigma_{\nu} J_{\nu} / h \nu \; \mathrm{d} \nu }{\int_{f_{\mathrm{ion}}}^{\infty} J_{\nu} / h \nu \; \mathrm{d} \nu },
\end{align}
where $J_{\nu}$ is the underlying spectrum of the source population.
In principle we could be more consistent in our choice of cell size by computing the effective length of the cell along the given direction $\Omega$ of the sweep, but we did not do so in order to keep this as simple as possible.

\begin{algorithm}
\caption{Chemistry update}\label{alg:chemistryUpdate}
\begin{algorithmic}[1]
\Procedure{Update}{$\Delta T$}
\State Remember initial state $U_{\mathrm{init}} \gets \bb{T, \xhii}$
\State Compute $T' \gets \Call{TemperatureUpdate}{\Delta t}$.
\State Compute $\xhii' \gets \Call{IonizationFractionUpdate}{\Delta t}$.
\If{$\abs{\frac{T' - T}{T}} > \epsilon$ \textbf{or} $\abs{\frac{\xhii' - \xhii}{\xhii}} > \epsilon$}
  \State $U \gets U_{\mathrm{init}}$
  \State \Call{Update}{$\Delta T / 2$}
  \State \Call{Update}{$\Delta T / 2$}
  \State \Return
\Else
  \State $T \gets T'$
  \State $\xhii \gets \xhii'$
\EndIf
\EndProcedure
\end{algorithmic}
\end{algorithm}

The basic chemistry update of a cell, given the incoming photon flux $F$ of photons above $\SI{13.6}{eV}$ proceeds as in Algorithm~\ref{alg:chemistryUpdate}. The temperature update is performed first, which means that the ionization fraction will be updated semi-implicitly using the new values of the temperature.

The temperature update is performed by solving the equation
\begin{align}
  \label{eq:temperatureEquation}
  \frac{\partial T}{\partial t} = \frac{{\protonMass \mu } \bb{\gamma - 1}}{\rho \kB} \Lambda
\end{align}
where $\protonMass$ is the mass of the proton, $\mu$ is the average mass of the particles in the gas in units of $\protonMass$, $\gamma$ is the adiabatic index, $\kB$ is the Boltzmann constant, $\rho$ is the mass density of the gas and $\Lambda$ is the total combined heating, and cooling term.
In our case, we assume that the gas consists only of hydrogen, so that $\mu = \frac{1}{1 + \xhii}$  where $\xhii$ is the hydrogen ionization fraction.

$\Lambda$ is given by a sum of the photo-heating term and the sum of all cooling processes,
\begin{align}
  \Lambda = H_{\mathrm{photo}}
  + \bb{\zeta(T) + \psi(T)} \electronDensity \neutralHydrogenDensity
  + \bb{\eta(T) + \Theta(T)} \electronDensity \ionizedHydrogenDensity
  + \bar{\omega}(T) \electronDensity,
\end{align}
where $H_{\mathrm{photo}}$ describes photo-heating, $\electronDensity$, $\neutralHydrogenDensity$ and $\ionizedHydrogenDensity$ are the (number-)density of electrons, neutral hydrogen and ionized hydrogen respectively and the other terms describe cooling due to collisional ionization $\zeta(T)$, collisional excitation $\psi(T)$, recombination $\eta(T)$, Bremsstrahlung $\Theta(T)$, and Compton cooling $\bar{\omega}(T)$.
We use the on-the-spot approximation in which we assume that every case A recombination (that is, recombination to the ground state) will emit a photon which is immediately re-absorbed by the surrounding neutral atoms so that it results in no additional recombination. Therefore $\eta(T)$ denotes the cooling rate of case B recombination only. 

In order to prevent instabilities related to the stiffness of the equations, we solve \equref{temperatureEquation} by updating the temperature via a semi-implicit formulation given by
\begin{align}
  \label{eq:temperatureUpdate}
  T^{t+\Delta t} = T^{t} + \frac{\mu \Lambda}{\frac{\rho \kB}{\bb{\gamma - 1} \protonMass \Delta t} - \Lambda'}.
\end{align}
Here, $\Lambda' = \frac{\partial \Lambda}{\partial T}$ is the derivative of the total heating rate with respect to temperature.

It should be noted, that while higher order methods could be used, we chose the above method for its simplicity and consistency with the approach developed by~\cite{rosdahlRamsesrtRadiationHydrodynamics2013}.
While this simple solver might suffer from excessive amounts of substepping if it were applied to a more complicated chemical network, we find that it performs well in our simple hydrogen-only network for the range of temperatures and ionization fractions that we are interested in.

The full expression for all the heating and cooling terms is given in \aref{subsweep-appendix-radchem}.

The equation describing the evolution of $\ionizedHydrogenDensity$ is given by
\begin{align}
  \label{eq:hydrogenEquation}
  \frac{\partial \ionizedHydrogenDensity}{\partial t} &= \neutralHydrogenDensity \bb{\beta(T) \electronDensity + \Gamma}- \alpha(T) \ionizedHydrogenDensity \electronDensity \\
  &= \hydrogenDensity \bb{(1 - \xhii) \bb{\beta(T) \electronDensity + \Gamma} - \xhii \alpha(T) \electronDensity} \\
  &= \hydrogenDensity \bb{(1 - \xhii) C - \xhii D}
\end{align}
where $\beta(T)$ is the electron collisional ionization rate, $\alpha(T)$ is the case-B recombination rate, $C$ and $D$ are the creation and destruction terms, $\Gamma$ is the photoionization rate, which is computed as $\sum_{i=1}^{n_{\rm faces}} \sum_{j=1}^{n_{\rm dir}} F_{i,j}$, where $n_{\rm faces}$ is the number of neighboring faces of the cell, $n_{\rm dir}$ is the number of discrete directions, and $F_{i, j}$ is the incoming photon flux from a given neighbor in the given direction, where $F_{i, j} = 0$ if the neighbor is downwind in the given direction.

Within a timestep, the ionization fraction is updated according to the semi-implicit formulation given by
\begin{align}
\xhii^{t + \Delta t} = \xhii^t + \Delta t \frac{C - \xhii^t (C + D)}{1 - J \Delta t},
\end{align}
where $J$ is given by
\begin{align}
    J = \frac{\partial C}{\partial \xhii} - (C + D) - \xhii \bb{\frac{\partial}{\partial \xhii} + \frac{\partial D}{\partial \xhii}}.
\end{align}

\subsection{Photon conservation\label{sec:photonConservation}}

A critical property of radiative transfer algorithms is their ability to conserve the amount of photons, meaning that the number of photons injected through sources should be equal to the number of photons that are either lost in photochemical reactions or leave the simulation box at the edge of the system (in a system without periodic boundary conditions).
The sweep algorithm described above is not manifestly photon-conserving, for two reasons.

The first reason is that the algorithm performs sub-timesteps. A higher-timestep cell will have its incoming photon flux determined by an earlier value of a neighbouring lower-timestep cell, which is not fully consistent with the value that the lower-timestep region has in its subsequent updates.

The second reason for photon non-conservation is that the sweep algorithm itself only uses an estimate of the number of photons which are going to be absorbed as a result of the photochemistry. This estimate is given by \eref{fluxEstimate} in terms of the photon flux. The actual number of photons which are going to be absorbed as a result of the photochemistry is only known after the chemistry step (after solving \eref{hydrogenEquation}) and the two values may differ. For example, the amount of photons which can be absorbed within any given cell should be limited by the number of atoms available for ionization. However, the initial estimate for the flux reduction in \eref{fluxEstimate} does not take this into account. As a result, the algorithm might severely overestimate the fraction of photons absorbed in the cell, resulting in lost photons.

An alternative to this split between the initial estimate and the actual computation would be to perform a chemistry update of the cell immediately upon encountering it during a partial sweep. However, there are multiple problems with this approach. The first is that it would increase the number of needed chemistry updates, since an update would have to be performed once per sweep direction instead of only once after all directional sweeps have finished. The second problem is that it introduces an artificial asymmetry - each sweep direction would encounter the cell in a different chemical state. To make matters worse, for the parallel sweep algorithm, this is not merely an artificial asymmetry but a non-deterministic one, since the order in which the sweep directions encounter the cell is dependent on the timings of the inter-processor communication and may differ from one simulation run to the next.
These reasons make performing a chemistry update for each sweep direction individually undesirable.

In order to address the latter cause of photon non-conservation, one approach is to limit the amount of photons that may be absorbed within any given cell by the amount of available atoms $N_{\mathrm{atoms}}$ that could be ionized by the photons. However, there is an immediate problem with this approach, namely that there are multiple sweep directions - each sweep would still be able to absorb as many photons as there are atoms, overestimating the number of absorbed photons by a factor of $N_{\mathrm{dir}}$. Alternatively, one could go even further and limit the maximum amount of absorbed photons to a fraction $N_{\mathrm{atoms}} / N_{\mathrm{dir}}$, to reduce this factor. However, this approach also has its drawback, since it may severely underestimate the amount of absorbed photons, leading to a widening of ionization fronts and increasing the velocity at which ionization fronts move through the medium.

We choose the former of these two approaches, that is, limiting the number of absorbed photons to $N_{\mathrm{atoms}}$ for each individual sweep direction. While this does not completely solve the problem of non-conservation, it can limit the amount of lost photons in extreme cases, without any of the unphysical effects that imposing a stronger limit would cause.

\section{Tests}\label{sec:tests}
\subsection{R-type expansion of an HII region\label{sec:rtype}}
\begin{figure}
    \centering
    \includegraphics[width=\columnwidth]{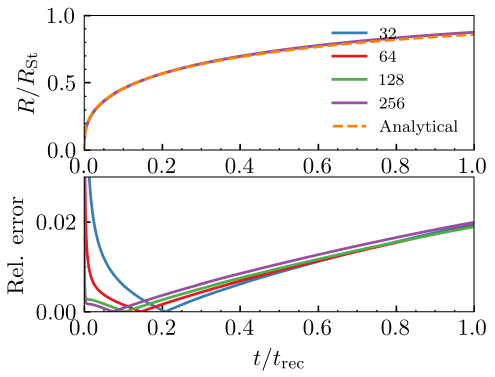}
    \caption{\label{fig:rtype_expansion}
      Results of the R-type expansion test.
      Top panel: The value of the radius of the ionized bubble at the center of the simulation box, in units of the Strömgren radius $\Rstroemgren$, plotted as a function of time (normalized by the recombination time $\trec$). Different lines represent different resolutions $32^3$ (blue), $64^3$ (red), $128^3$ (green), $256^3$ (purple) with the orange, dashed line representing the analytical prediction given by \eref{stroemgrenAnalytical}.
      Bottom panel: The relative error $\abs{R(t) - \Rr(t)} / \Rr(t)$ between the analytical prediction $R(t)$ and the numerical results $\Rr(t)$ as a function of $t / \trec$.
    }
\end{figure}

Here we study the R-type expansion of an HII region. \citet{stromgrenPhysicalStateInterstellar1939} showed that a point source in a medium consisting of hydrogen with uniform density will eventually create a spherical HII region with radius $R_{\text{st}} = \bb{\frac{3 \dot{N}}{4 \pi \alpha_{\text{B}}(T) \electronDensity^2}}^{1/3}$ where $\dot{N}$ is the rate of ionizing photons emitted from the source, $\alpha_{\text{B}}$ is the temperature-dependent case B recombination coefficient, and $\electronDensity$ is the electron density.
If we assume that the gas inside the ionized region is fully ionized then $\electronDensity$ is equal to the number density of H nuclei, $\hydrogenDensity$, and we find that the time evolution of the system is given by
\begin{align}
  \label{eq:stroemgrenAnalytical}
  R(t) = R_{St} \bb{1 - e^{-t / t_{\text{rec}}}},
\end{align}
where the recombination time $t_{\text{rec}}$ is given by $t_{\text{rec}} = \frac{1}{\hydrogenDensity \alpha_{\text{B}}(T)}$.
This test is set up in exactly the same way as described by~\citet{jauraSPRAICouplingRadiative2018} and~\citet{baczynskiFerventChemistrycoupledIonizing2015}, as well as the corresponding test in \citet{peterSweepMethodRadiative2023}. We use a cubic simulation box with $L = \SI{12.8}{\kilo \parsec}$ filled with hydrogen with a homogeneous density of $\hydrogenDensity = \SI{1e-3}{\per \cubic \centi \meter}$ which is assumed to be fully neutral in the beginning. In the center of the box, we place a single source which emits ionizing photons (of energy $E > \SI{13.6}{eV}$ ) at a rate of $\dot{N} = \SI{1e49}{\per \second}$. During the test, we disable photo-heating and all cooling terms, keep the temperature at $T = \SI{100}{K}$ everywhere, and fix the case B recombination coefficient to $\alpha_{\mathrm{B}} = \SI{2.59e-13}{cm^3 \, s^{-1}}$.
We perform this test for a number of different resolutions, $N_{\rm cell}=32^3,64^3,128^3,256^3$.
The top panel of \figref{rtype_expansion} shows the numerical result for the radius of the ionized bubble as a function of time, compared to the analytical expression given by \equref{stroemgrenAnalytical}.
In the bottom panel, the relative error between the simulation result and the analytical prediction is shown.
The radius of the ionized bubble is defined as the radius $r$ at which a small spherical shell with radius $r$ has an average ionization of $\overline{x_{\mathrm{HII}}} = 0.5$.
For more details regarding the computation of this value, see \citet{peterSweepMethodRadiative2023}.

The runs at all resolutions follow the analytical prediction closely with relative errors of below $2 \%$ for all values the initial phase of the expansion where $t < 0.1 t_{\mathrm{rec}}$.
It should be noted that the general trend is that the error increases with increasing resolution, a result that we have also seen in the {\sc Sweep} implementation of {\sc Arepo}.
In the following, we give two possible reasons for this counterintuitive result. 
First of all, we note that the analytical prediction assumes a perfectly sharp boundary, something that is clearly not the case in the numerical solution to the problem. The thickness $w$ of the ionization front (defined as the distance between the point at which the average ionized fraction is $\SI{10}{\percent}$ and the point at which it is $\SI{90}{\percent}$) that we find in our simulation is roughly $w = \SI{1.0}{\kilo \parsec}$ at the lowest resolution ($32^3$ particles) and decreases slightly to about $w = \SI{0.7}{\kilo \parsec}$ at the highest resolution ($256^3$ particles). We note that this is very close to the analytical prediction of $w = \SI{0.74}{\kilo \parsec}$ for the same test given in \citet{ilievCosmologicalRadiativeTransfer2006}.
This means that the value of the radius depends strongly on its definition, so that small deviations in the radius are not necessarily meaningful.

The second reason for the slight reduction in accuracy with increasing resolution might be related to the issue of photon non-conservation discussed in \sref{photonConservation}. The effects explained there might reasonably lead to something like the observed result. In order to understand this better, we study the quantitative effect of photon-non-conservation in \secref{photonConservationTest}.

\subsection{Test of periodic boundary conditions\label{sec:periodic}}
In order to test the behavior of the algorithm in setups with periodic boundary conditions, we perform a test similar to the one in \citet{peterSweepMethodRadiative2023}. In this test we perform a R-type expansion of an HII region as in the previous section (\secref{rtype}). The difference between the two simulations is that in this test, rather than placing the source in the center of the box, we place it right at the boundary in the $x$-direction at position $\bm{r} = (0, 6.4, 6.4) \; \mathrm{kpc}$. A slice through the box illustrating the setup of the test is shown in \fref{periodic_rtype_slice}.

\begin{figure}
    \centering
    \includegraphics[width=\columnwidth]{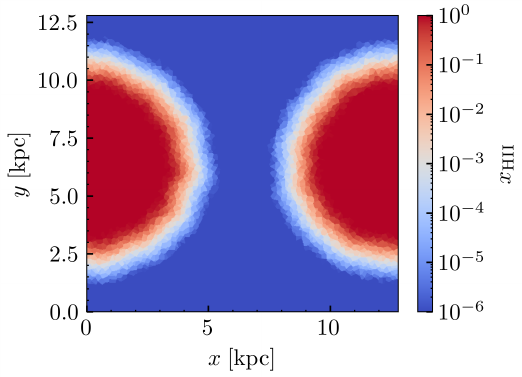}
    \caption{\label{fig:periodic_rtype_slice}
      Slice through the box in the plane $z = \SI{6.4}{\kilo\parsec}$ at t=\SI{20}{Myr}. The color shows different values of the ionization fraction with blue being neutral and red being ionized.
    }
\end{figure}

In the simulation there is no cell at exactly that position, so the source term will be introduced into a cell that is slightly to the right (at positive $x$) of $\bm{r}$, namely at $\bm{r} = (\epsilon, 6.4, 6.4) \; \mathrm{kpc}$, where $\epsilon > 0$ is small.
Since the source is placed so close to the edge of the simulation box at $x=0$, any photons originating at the source with a direction to the left must first pass through the (periodic) boundary before they re-enter from the other side and begin having an effect on the gas.
Since the only symmetry breaking element in this setup is the simulation box itself, an accurate solver should produce a reasonably symmetric result, up to the precision determined by the resolution of the simulation.
If photons exiting the boundary are not re-introduced on the other side consistently, we will notice a lack of ionization in the right side of the box, compared to the left side.

In order to quantify how well our solver deals with periodic boundary conditions, we compute the asymmetry $a$, defined as the relative difference between the average ionization fraction in the left side of the box and the right side of the box given by
\begin{align}
  \label{eq:asymmetry}
  a = \abs{\frac{\bar{x}_{\text{HII, left}} - \bar{x}_{\text{HII, left}}}{\bar{x}_{\text{HII, left}} + \bar{x}_{\text{HII, left}}}},
\end{align}
where the (volume-)averaged ionization fractions $\bar{x}_{\text{HII, left}}$ and $\bar{x}_{\text{HII, right}}$ are defined as
$$\bar{x}_{\text{HII, left}} = f(\epsilon, L/2 + \epsilon)$$
and
$$\bar{x}_{\text{HII, right}} = f(0, \epsilon) + f(L/2+\epsilon, L),$$
with 
$$f(x_1, x_2) = \frac{2}{V} \int_{x_1}^{x_2} \mathrm{d}x \int_{-L/2}^{L/2} \mathrm{d}y \int_{-L/2}^{L/2} \mathrm{d}z\; x_{\text{HII}}.$$
That is, the two averages are computed over the left- and right- halves of the simulation box as seen from the source at $\epsilon$, which corresponds to the left- and right- halves of the box except for the small $\epsilon$-sized sliver on the left.

From our implementation of periodic boundary conditions, we can expect that smaller timesteps will achieve more accurate results than larger timesteps, since the initial estimate of the photon fluxes is given simply by the fluxes from the previous timestep: if the timestep itself is small, the prediction will be more accurate.
However, the main goal of this test is not just to test the behavior of the solver with respect to the timestep, but to check whether allowing the solver to perform sub-timesteps has a positive effect on the accuracy.
In order to test this, we perform the simulation setup described above for a variety of timesteps and different numbers of sub-timestep levels $n$ and compute the periodic asymmetry given by \eref{asymmetry}.

\begin{figure}
    \centering
    \includegraphics[width=\columnwidth]{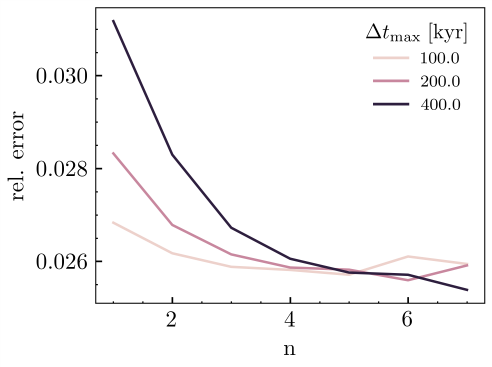}
    \caption{\label{fig:periodic_asymmetry}
      Asymmetry of the average ionization fraction. The asymmetry (see \eref{asymmetry}) is shown as a function of the number of timestep levels used for the test. The different lines depict different values for the maximum timestep used.
    }
\end{figure}

In \fref{periodic_asymmetry} we show the asymmetry $a$ as a function of the number of timestep levels $n$ for different values of the maximum timestep $\Delta t_{\text{max}}$.
Initially, it should be noted that the asymmetry is already quite small, with values below $a < \SI{0.4}{\percent}$ even at a timestep of $\Delta t_{\mathrm{max}} = \SI{400}{\kilo\year}$, which corresponds to $\Delta t_{\mathrm{max}} \approx \frac{3}{1000} t_{\mathrm{rec}}$ (the recombination time in this test is the same as in \sref{rtype}).
Despite this, we still see the expected overall trend, which is that the asymmetry decreases as the timestep decreases, in approximately linear fashion.
Moreover, allowing sub-timestepping to use more levels also decreases the asymmetry, with a clear downwards trend for $n \leq 5$.
At $n = 6$ and $n = 7$ the asymmetry increases temporarily.
While this may initially seem worrying, the magnitude of the asymmetry is already below the values that we can reasonably expect to resolve given the relatively low resolution of this test.
We also note that the numerical result of this test is quite strongly dependent on the exact value of $\epsilon$ - while changing it slightly does not change the overall trend, it does change the absolute values of the asymmetry, especially for values below $a < 0.001$.

\subsection{Shadowing behavior behind an overdense clump\label{sec:shadowing}}
\begin{figure*}
    \centering
    \includegraphics[width=1.0\textwidth]{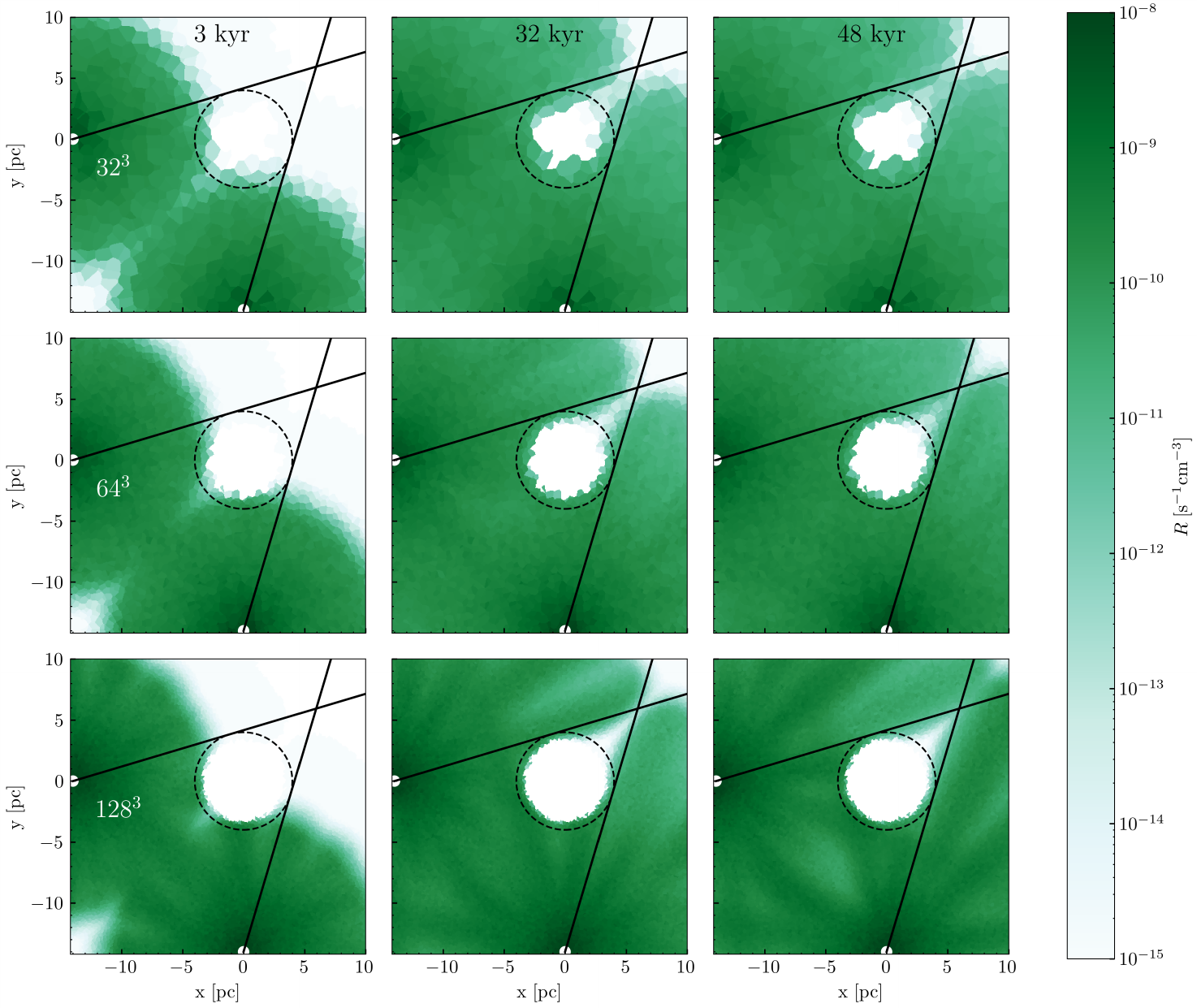}
    \caption{\label{fig:shadowing2}Photon rate $R$ in a slice through the $z=0$-plane of the simulation box.
      First row: $32^3$ particles,
      Second row: $64^3$ particles,
      Third row: $128^3$.
      First column: $t = \SI{3.0}{\kilo\year}$,
      second column: $t = \SI{32}{\kilo\year}$,
      third column: $t = \SI{48}{\kilo\year}$,
      The black dashed circle represents the overdense clump.
      White solid circles represent the position of the sources.
      The black dashed lines delineate the shape of an ideal shadow behind the clump.
    }
\end{figure*}
Due to the radial symmetry of the R-type expansion shown in \secref{rtype}, it does not probe the directional properties of the radiative transfer itself very well. To do so, we perform the following test which is also identical to the setup in~\citet{jauraSPRAICouplingRadiative2018} and~\citet{baczynskiFerventChemistrycoupledIonizing2015}.
We study the formation of a shadow behind an overdense clump. We first set up a box of length $L = \SI{32}{\parsec}$. The box is filled with hydrogen with a spatially varying density with
\begin{align}
  \hydrogenDensity(\bm{x}) =
  \begin{cases}  
    \SI{1000}{\per \cubic \centi \meter} &\text{where } |x| < \SI{4}{\parsec},   \\
    \SI{1}{\per \cubic \centi \meter} &\text{everywhere else}.
  \end{cases}
\end{align}
We place two point sources at $\bm{r}_1 = (-14, 0, 0) \; \mathrm{pc}$ and $\bm{r}_2 = (0, -14, 0) \; \mathrm{pc}$, which emit photons at a rate of $\dot{N} = \SI{1.61e48}{\per\second}$.
An analysis of this test, which includes hydrodynamics and discusses the temperature, pressure, and density response has been performed by \citet{jauraSPRAICouplingRadiative2018}.

In \figref{shadowing2}, we show the rate of photons passing through the cell in units of $\SI{}{cm^{-3} s^{-1}}$ in a slice through the simulation box along the x-y plane for different times (columns) and resolutions (rows).
We find that the algorithm will correctly form a shadow behind the overdense clump.
However, due to numerical diffusion, the shadow does not follow the theoretically expected form exactly.
Because of this, the region behind the overdense clump will slowly begin to ionize.
In order to quantify the shadowing behavior we calculate the mass-averaged fraction of ionized hydrogen in the volume of the shadow.
The volume is given by the intersection of two (infinitely extended) cones, with their tips at $\bm{r}_1$ and $\bm{r}_2$ respectively and their base determined by the great circle lying in the overdense clump. The overdense clump itself is excluded from the volume.
In the 2D slice shown in \fref{shadowing2}, this volume $V_{\rm S}$ corresponds to the area between the black dashed circle and the black lines.
The average ionization fraction in the shadow region $\bar{x_{\rm HII}}$ is given by
\begin{align}
  \label{eq:volumeIntegral2}
  \overline{x_{\rm H}} = \frac{\int_{V_{\rm S}} x_{\rm H}(\bm{r}) \rho(\bm{r})  \mathrm{d}V}{\int_{V_{\rm S}}  \rho(\bm{r})  \mathrm{d}V},
\end{align}
where $x_{\rm H}(\bm{r})$ is the abundance of ionized hydrogen at position $\bm{r}$ and $\rho(\bm{r})$ denotes the mass density at position $\bm{r}$.

\begin{figure}
  \centering
    \includegraphics[width=\columnwidth]{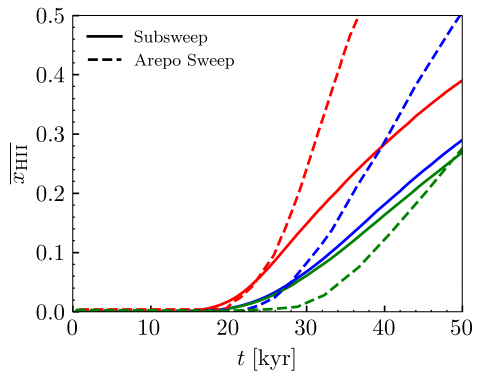}
    \caption{\label{fig:shadowingVolume2}
      Results of the test of the shadowing behavior of the Sweep method.
      The average ionized hydrogen abundance $\overline{x_{\rm HII}}$ (see \eref{volumeIntegral2}) in the shadow volume as a function of time for {\sc Subsweep} (solid lines) and the original {\sc Sweep} implementation in {\sc Arepo} (dashed lines) for three different resolutions:
      $128^3$ (green),
      $64^3$ (blue) and
      $32^3$ (red),
    }
\end{figure}

In \figref{shadowingVolume2}, $\overline{x_{\rm H}}$ is shown as a function of time. 
The ionization fraction begins to increase at $t \approx \SI{20}{\kilo \year}$, demonstrating that the sweep algorithm does not form a perfect shadow.
The shadowing behavior improves going from lower resolution to higher resolution.
This is in line with the explanation that the protrusion of the ionization front into the shadow is due to numerical diffusion, since higher resolutions decrease the effect of numerical diffusion.
We also find that for low resolutions ($32^3$, $64^3$), {\sc Subsweep} forms a better shadow than the {\sc Sweep}, especially at late times, while {\sc Sweep} performs better in the high resolution case $128^3$.
It might be surprising that there are different results at all, considering the fact that the two implementations use the same algorithm for the radiation transport.
This is due to the fact that the chemistry updates are performed differently.
The {\sc Arepo} implementation, {\sc Sweep} will perform a radiation chemistry update for each time a directional sweep encounters a cell.
In {\sc Subsweep}, cell abundances are fixed until the end of the sub-timestep and therefore remain the same for each directional sweep.
This can result in different behavior at the ionization front.

\subsection{1D r-type expansion\label{sec:oneDExpansion}}
In order to test the behavior of the substepping algorithm, we perform a test in which we study the expansion of an ionization front in a one dimensional box filled with hydrogen of uniform number density $n = \SI{1e-4}{\per \cubic \centi \meter}$. The medium extends from $0$ to $L$. The gas is kept at a constant temperature $T = \SI{100}{\kelvin}$. A source emitting a constant flux of ionizing photons of $\SI{1e5}{\per \square \centi \meter \per \second}$ in the direction towards the right is placed at $x = 0$.
The time evolution of this system is characterized by the formation of an ionized region of all cells with $0 < x < D(t)$ where $D(t)$ is the size of this ionized region as a function of time and given by $D(t) = D_{\mathrm{st,1d}} \bb{1 - e^{-t / t_{\text{rec,1d}}}}$. Here $D_{\text{st,1d}}  = \frac{F}{\alpha_{\text{B}} n^2}$ is the one-dimensional Strömgren length, and $t_{\text{rec,1d}} = \frac{1}{\alpha_{\text{B}} n}$ is the recombination time.

For the numerical simulation of the system, we divide the interval into $N$ equidistant cells with width $\frac{L}{N}$ along the line. The leftmost cell contains the source. In this test, we only perform sweeps in one direction (pointing to the right).
In practice, the ionization front will not be infinitely thin but extend over several cells.
If $N$ is large enough such that the ionization front is well-resolved, we can expect the analytical expression for the total ionized volume fraction $x_{\text{analytical}} = \frac{L(t)}{L}$ to accurately predict the numerical result $x_{\text{numerical}}$ , so that we can define a simulation to have converged to the right result if
\begin{align}
  \label{eq:convergence1d}
  \abs{x_{\text{numerical}} - x_{\text{analytical}}} < \epsilon,
\end{align}
where $\epsilon > 0$ is the error tolerance which we choose to be $\epsilon = 1 \%$.
We want to study the convergence behavior of the sweep algorithm in this system. In order to do so, we perform runs with different values for the maximum number of allowed timestep levels $n$, namely $n = \numlist{1;2;3;4;5;6}$ as well as different numbers of cells $N = \numlist{160;320;640;1280;2560;5120;10240}$.
For each set of values of $n$ and $N$, our goal is to find the highest value of the timestep $\Delta t$ for which the simulation reaches the correct result, that is where \eref{convergence1d} holds.

\begin{figure}
    \centering
    \includegraphics[width=\columnwidth]{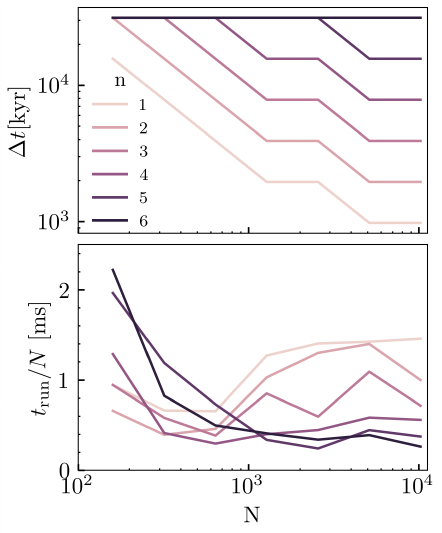}
    \caption{\label{fig:convergence1d}
        Overview of the convergence behavior of the Subsweep method. 
        Top panel: The largest converging timestep $\Delta t$ as a function of the number of particles $N$. The different lines represent runs with different number of allowed timestep levels $n$. Bottom panel: The total runtime $t_{\mathrm{run}}$ divided by the number of particles $N$.}
\end{figure}

In \figref{convergence1d}, we show the maximum converging timestep $\DeltaT$ as a function of $N$ for each value of $n$.
For low values of $N$, adding more timestep levels does not result in a meaningfully different result. However, as $N$ increases, the separation between runs at different $n$ becomes clear, with higher values of $n$ increasing the highest possible converging timestep.
This result clearly demonstrates that using substepping allows us to use larger timesteps while still converging to the physically correct result.
In fact, each additional substepping level allows us to increase the maximum timestep $\DeltaT$ by a factor of two, which is the expected outcome.

It should be noted that this is a trivial implication if all cells were kept on the lowest timestep during the simulation, since in this case the subsweeping algorithm reduces to performing $2^{n-1}$ sweeps of the full system with a timestep of $2^{-(n-1)} \DeltaT$ each.
In order to demonstrate that this is not the case and that we have gained something from the subsweeping, the bottom panel of \fref{convergence1d} shows the total runtime $t_{\mathrm{run}}$ of the corresponding simulation in the top panel, divided by the number of particles $n$.
This clearly shows that, for large $N$, using more timestep levels alleviates the need to use a low, global timestep which in turn reduces the simulation time significantly, while still producing physically accurate results.
For small $N$, the substepping does not improve performance and at times will even decrease performance.
One possible explanation for this could be that most if not all of the cells in the simulations are at a very low timestep.
While having a large number of timestep levels will not change the numerical result of the simulation, it can decrease performance due to the additional computational overhead of communicating the levels of each of the cells multiple times for each timestep.

\subsection{Photon conservation\label{sec:photonConservationTest}}

The goal of this section is to quantify the fraction of photons that are lost as a result of the non-conservation of photons in our solver. We perform all of the following tests by using the simulation setup described in \sref{shadowing}. In order to compute the fraction of lost photons, we use a special version of the chemistry solver in which all heating and recombination processes are effectively disabled, leaving ionization of hydrogen as the only remaining process. In order to keep the impact that this change to the solver has on the result of the simulation to a minimum, we only perform a single, full sweep timestep, keeping track of the amount of photons injected $N_{\mathrm{injected}} = L \Delta t_{\mathrm{max}}$, where $L$ is the total luminosity, the amount of photons lost at the boundaries of the simulation $N_{\mathrm{boundary}}$, as well as the amount of photons absorbed as a result of an ionization process in a cell $N_{\mathrm{absorbed}}$. Using these three values, we can simply compute the fraction $\xi$ of photons that are being lost as a result of the solver as

\begin{align}
\label{eq:photonLossFraction}
\xi = \frac{N_{\mathrm{injected}} - (N_{\mathrm{absorbed}} + N_{\mathrm{boundary}})}{N_{\mathrm{injected}}}.
\end{align}

In order to understand the impact that different time-stepping parameters and spatial resolutions have on photon non-conservation, we run a number of simulations with different resolutions, $N_{\rm cell}=32^3,64^3,128^3,256^3$, different timesteps $\Delta t_{\mathrm{max}}$ and a different maximum number of timestep levels $n=1,2,3,4,5,6,7$. Additionally, we perform each simulation both with and without the absorption limiter described in \secref{photonConservation}.

\begin{figure}
    \centering
    \includegraphics[width=\columnwidth]{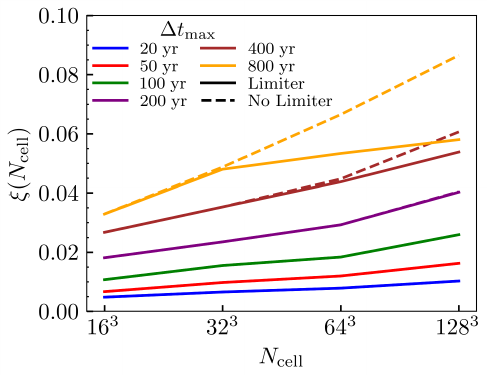}
    \caption{\label{fig:conservation_resolution}Fraction of lost photons $\xi$ as a function of the number of particles $N_{\rm cell}$. Different colors show results for different maximum timesteps $\Delta t_{\rm max}$. Solid lines show the result with the absorption limiter enabled, dashed lines show results without the absorption limiter.}
\end{figure}

In \figref{conservation_resolution}, we show the fraction of lost photons $\xi$ as a function of the number of cells $N_{\mathrm{cell}}$ for different timesteps $\Delta t_{\rm max}$. All simulations shown here were performed without substepping (with $n = 1$). The first thing to note is that the loss ratio remains below \SI{10}{\percent} in all simulations and well below it in most of the cases. This result also shows that the loss ratio decreases with decreasing timestep. This is an intuitive result, since the difference between the estimate (given by \eref{fluxEstimate}) and the actual chemistry update (given by \eref{hydrogenEquation}) vanishes with sufficiently small timesteps. The result also demonstrates that the photon loss ratio increases with increasing spatial resolution. In a way, this is a similar result to the trend we observed before, given the fact that the timestep required for accurate integration of the system will decrease with decreasing cell size. Moreover, we find that if the timestep and the spatial resolution are comparatively high (if the timestep integration is sufficiently inaccurate), enabling the absorption limiter will strongly reduce the amount of lost photons.

\begin{figure}
    \centering
    \includegraphics[width=\columnwidth]{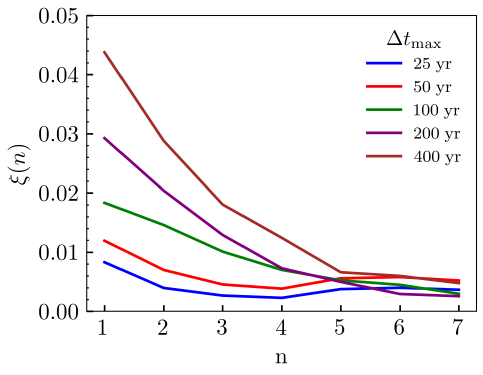}
    \caption{\label{fig:conservation_n}Fraction of lost photons $\xi$ as a function of the number of allowed timestep levels $n$ for the simulation with $N_{\mathrm{cell}} = 64^3$. Different colors show results for different maximum timesteps $\Delta t_{\rm max}$.}
\end{figure}

In \figref{conservation_n}, we show the photon loss fraction $\xi$ as a function of the number of allowed timestep levels $n$ for the simulation setup with $N_{\mathrm{cell}} = 64^3$. The dominating trend is that increasing the number of timestep levels decreases the photon loss fraction. This is a striking result, given the fact that sub-timestepping is one of the reasons for photon non-conservation. However, this result can be well explained by the trend shown in \fref{conservation_resolution}, since increasing the amount of timestep levels results in an effectively decreased timestep, increasing the overall accuracy of the time integration and therefore reducing the amount of lost photons. However, for the small timesteps $\Delta t_{\rm max} = \SI{25}{\year}$ and $\Delta t_{\rm max} = \SI{50}{\year}$, we also see that increasing the number of timestep levels beyond a certain point ($n = 4$ here) has the opposite effect of slightly increasing the loss fraction. Given that these are the simulations which already perform a reasonably accurate time-integration, even at $n=1$, we believe that this increase is in fact explained by the lost photons due to sub-timestepping which end up dominating the small improvements that further increasing $n$ has on the time-integration.

\subsection{Tests of the radiation chemistry\label{sec:chemistryTests}}
In order to test the radiation chemistry, we perform a series of tests that follow those performed in \citet{rosdahlRamsesrtRadiationHydrodynamics2013} as closely as possible.
The difference between the two setups is that our radiation chemistry solver does not incorporate helium, so some of the results can look different.
However, we still expect the results to be qualitatively very similar.
For all these tests, we take a single cell which we initialize with a given temperature, density, fraction of ionized hydrogen, and ionizing flux.
We perform tests with all possible combinations of 
densities between $\numlist{1e-8; 1e-6; 1e-4; 1e-2; 1e0; 1e2}~\si{\per \cubic \centi \meter}$,
initial ionized hydrogen fractions between $\numlist{0; 0.2; 0.5; 0.8; 1.0}$,
initial temperatures between $\numlist{1e3; 1.6e4; 1.8e5; 3e6; 1e8} \si{K}$
and either zero ionizing photon flux, or a ionizing photon flux of $\SI{1e5}{\per \second \per \centi \meter \squared}$.
For all of these configurations, we perform normal time evolution tests in which we let the system evolve freely.
For each test, we let the system evolve for a total time of $t = \SI{10}{\giga\year}$, which corresponds to $t \approx 8 t_{\mathrm{rec}}$ at $\SI{1e-4}{\per \cubic \centi \meter}$ and to $t \approx 10^{-3} \; t_{\mathrm{rec}}$ at $\SI{1e-8}{\per \cubic \centi \meter}$.

In \fref{chemistry_evolution_xhii} we show the hydrogen ionization fraction as a function of time for a subset of the parameters. 
We find that most configurations eventually converge, but both the limit and the convergence time vary drastically between the different configuration.

In the case of zero ionizing flux (top panel of \fref{chemistry_evolution_xhii}), the configurations that do not converge are those with very low densities ($n \leq \SI{1e-6}{\per\cubic\cm}$).
There, recombination rates and collisional ionization rates are extremely low due to the $n^2$ dependence.
The lack of ionizing flux results in zero photoionization, so that the ionization fraction in these cells remains constant over extremely long times.
In all other cases, we find that the ionization fraction always converges either to $0$ or $1$.
At high enough densities ($n \geq \SI{1e-2}{\per\cubic\cm}$), the cell always ends up fully neutral after long enough times.
If the initial temperature is high enough and the density low enough, the cell can become fully ionized, although it might eventually become neutral again on extremely long timescales, but this would require running the tests for even longer timescales.
Since timescales of longer than $\SI{10}{\giga\year}$ are not relevant for our application we will refrain from running these tests for even longer.

If an ionizing flux of $F = \SI{1e5}{\per \second \per \square \cm}$ is present (bottom panel of \fref{chemistry_evolution_xhii}), virtually all configurations at low densities are immediately ionized.
Only at comparatively high densities ($n \geq \SI{1e0}{\per\cubic\cm}$) can recombination dominate such that cells reach an equilibrium value (of approximately $x_{\mathrm{HII}} = 0.4$ for $n = \SI{1e0}{\per\cubic\cm}$ and very close to $x_{\mathrm{HII}} = 0.0$ for $n = \SI{1e2}{\per\cubic\cm}$).

\begin{figure*}
  \begin{minipage}[c]{0.50\textwidth}
    \centering
    $F = 0$ \\
    \includegraphics[width=1.0\textwidth]{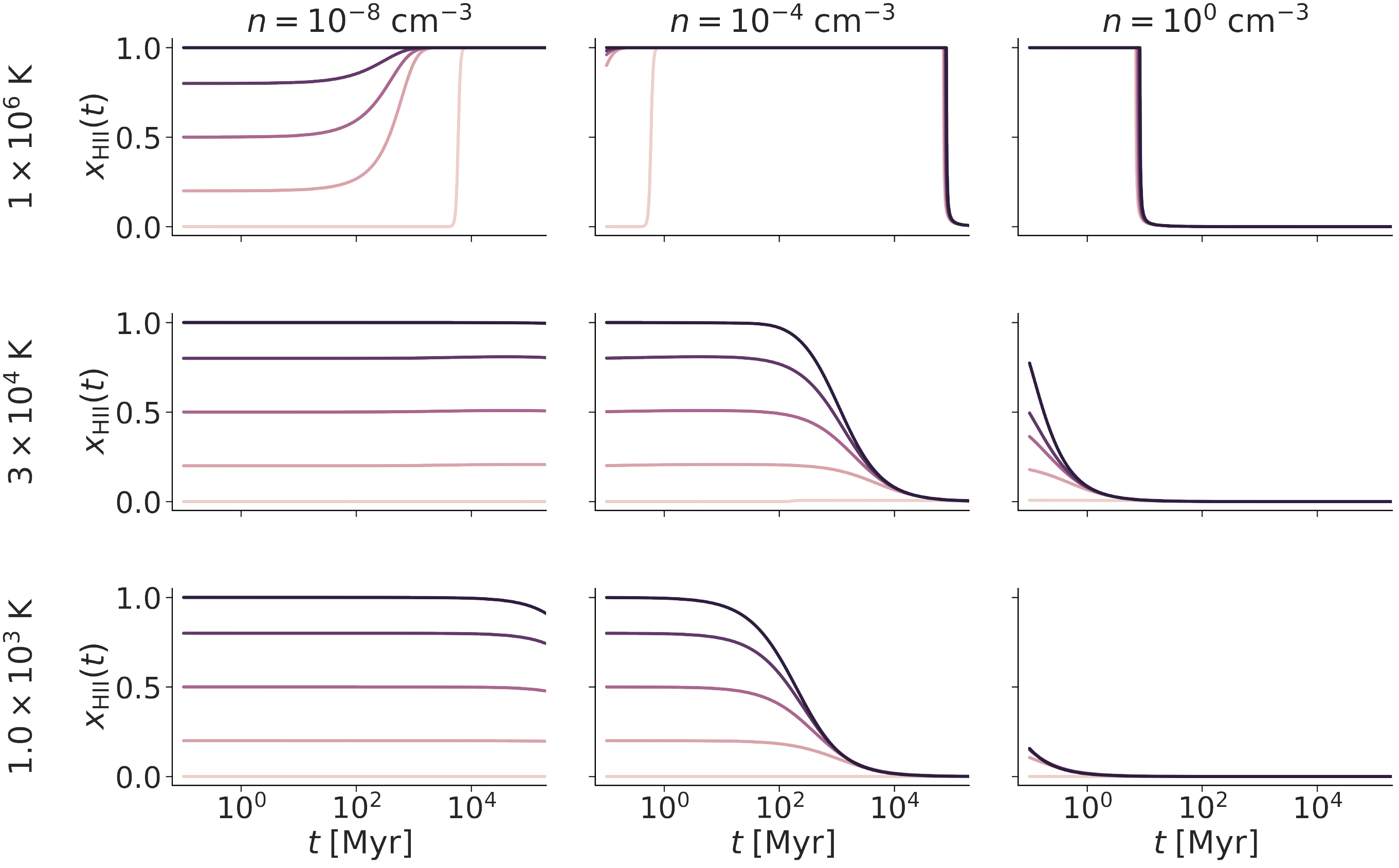} \\
    $F = \SI{1e5}{\per \second \per \square \cm}$ \\
    \includegraphics[width=1.0\textwidth]{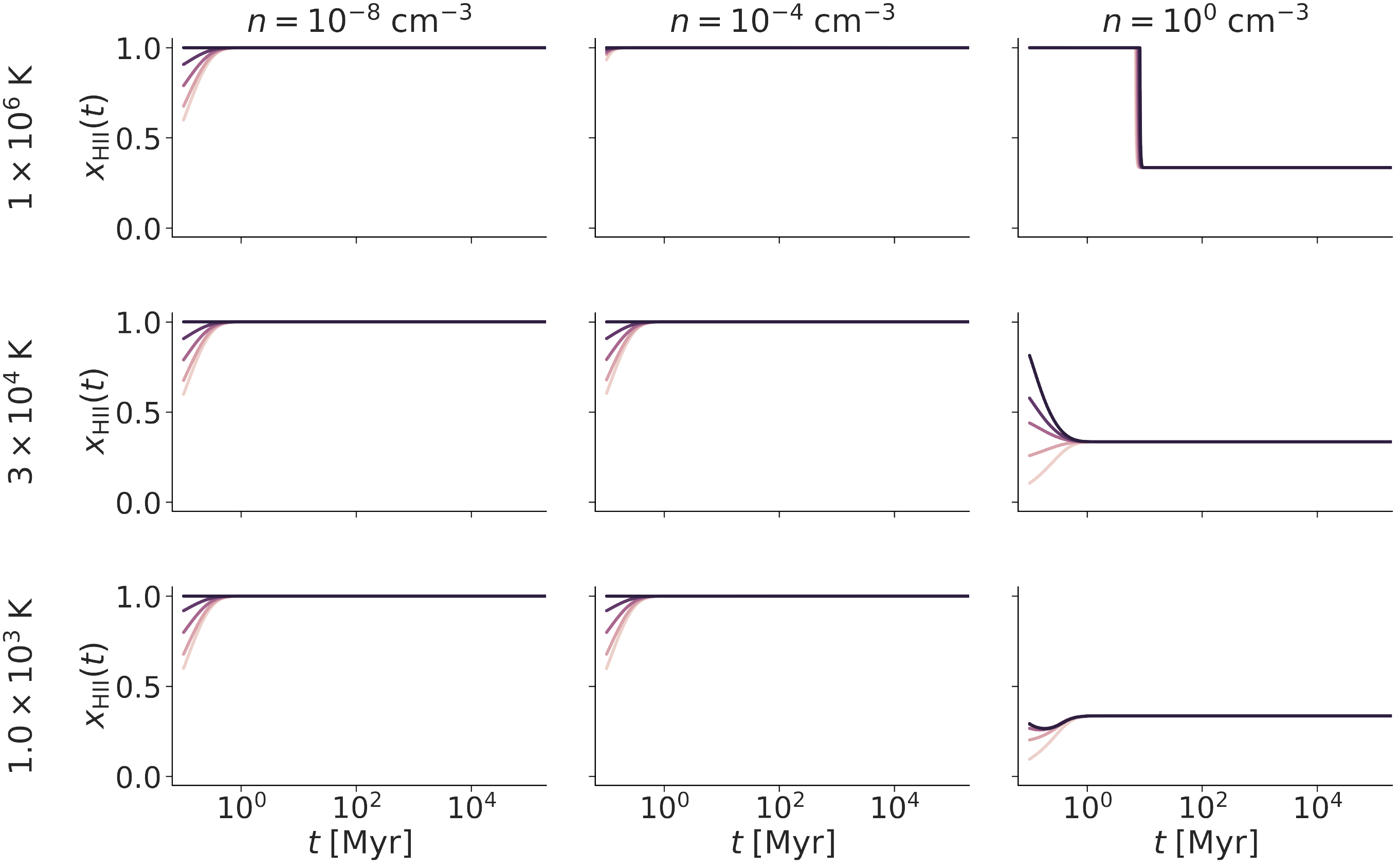} \\
  \end{minipage}\hfill
  \begin{minipage}[c]{0.36\textwidth}
    \caption{
       \label{fig:chemistry_evolution_xhii}Time evolution of the ionized hydrogen ionization fraction for different values of the density in the cell (columns), different initial temperatures (rows), and different values of the initial ionized fraction (line colors).
      Top panel: no ionizing flux.
      Bottom panel: with an ionizing flux of \SI{1e5}{\per \second \per \square \cm}
    }
  \end{minipage}
\end{figure*}

In \fref{chemistry_evolution_temp} we show the temperature as a function of time for a subset of the parameters.
As above, most configurations eventually converge, with the exception of zero ionizing flux in the case of very low densities.
In the absence of an ionizing flux (top panel of \fref{chemistry_evolution_temp}), cells never heat and cool down on timescales determined by their densities.
For the case of an ionizing flux of $F = \SI{1e5}{\per \second \per \square \cm}$ (bottom panel of \fref{chemistry_evolution_temp}), equilibrium temperatures are on the order of $\SI{1e4}{\kelvin}$, with convergence time being strongly affected by the cell density.

\begin{figure*}
  \begin{minipage}[c]{0.50\textwidth}
    \centering
    $F = 0$ \\
    \includegraphics[width=1.0\textwidth]{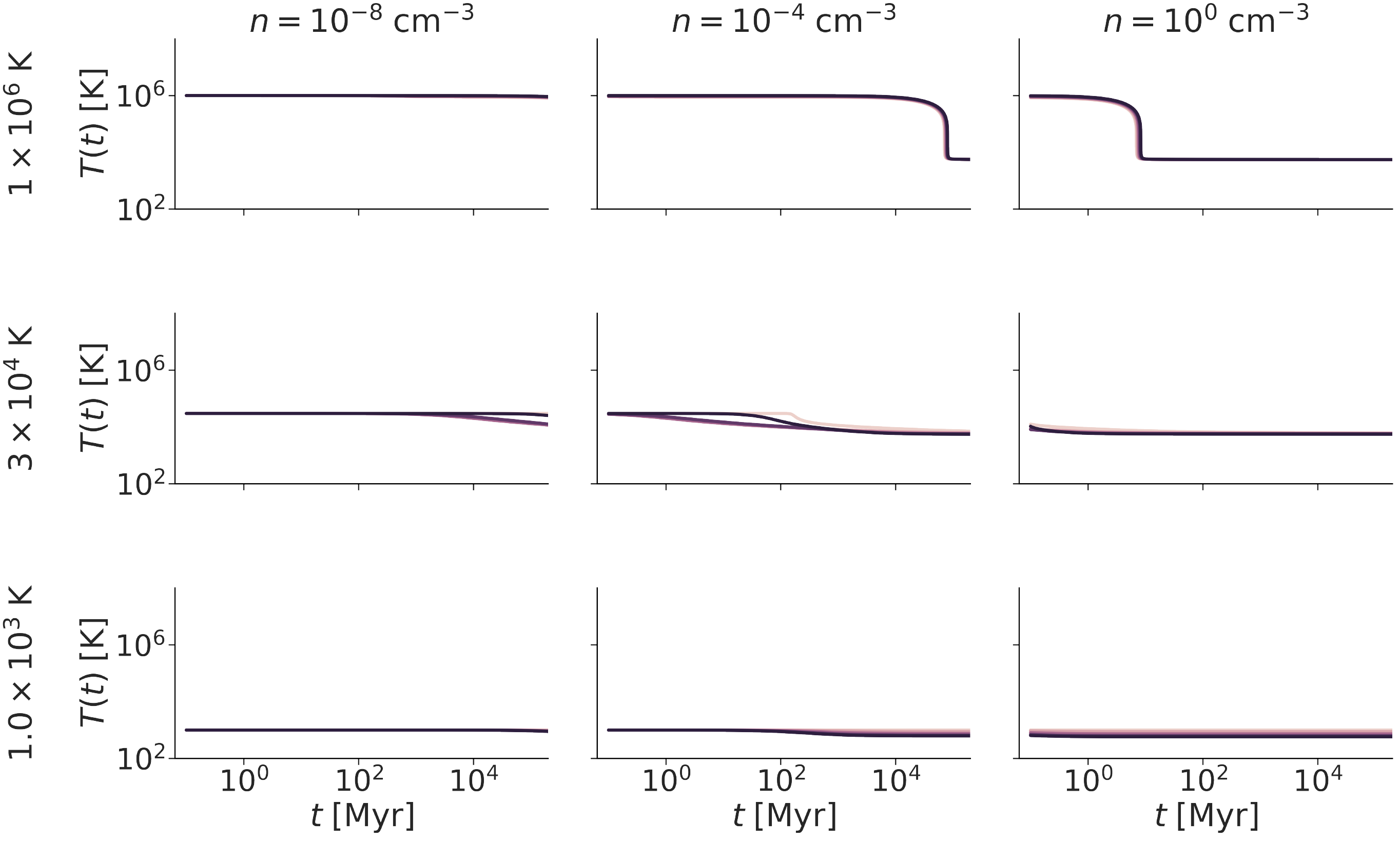} \\
    \centering
    $F = \SI{1e5}{\per \second \per \square \cm}$ \\
    \includegraphics[width=1.0\textwidth]{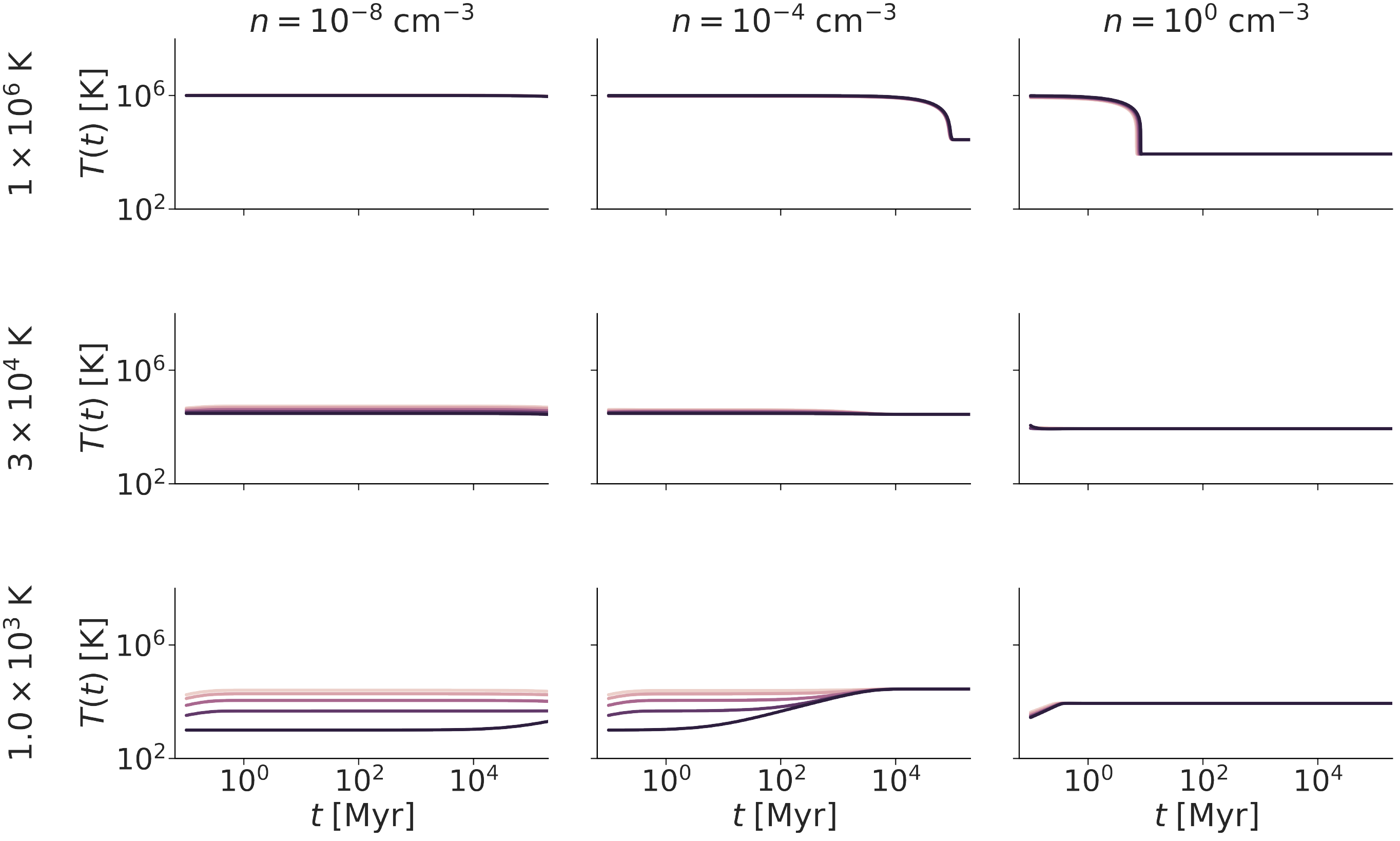} \\
  \end{minipage}\hfill
  \begin{minipage}[c]{0.36\textwidth}
    \caption{
       \label{fig:chemistry_evolution_temp}Time evolution of the cell temperature for different values of the density in the cell (columns), different initial temperatures (rows), and different values of the initial ionized fraction (line colors).
      Top panel: no ionizing flux.
      Bottom panel: with an ionizing flux of \SI{1e5}{\per \second \per \square \cm}
    }
  \end{minipage}
\end{figure*}

\section{Conclusion}\label{sec:conclusion}
In this paper, we introduced the radiative transfer postprocessing code {\sc Subsweep}, a standalone code that takes input from astrophysical hydrodynamics simulation codes (currently only {\sc Arepo} is supported, but extensions are easily possible) and performs radiative transfer on the input.
We discussed the choice and implementation of the domain decomposition as well as the algorithm for the Voronoi grid / Delaunay triangulation employed to efficiently construct a Voronoi grid in parallel.

We briefly discussed the sweep method and its original implementation {\sc Sweep}, which is a discrete ordinate method. It gives the exact solution to the scattering-less radiative transfer equation in a single pass over the grid in order to introduce the main feature of {\sc Subsweep} - the extension of the sweep method to incorporate sub-timestep sweeps in order to solve the coupled system of radiative transfer and radiation chemistry more efficiently.
We described how the substepping is implemented by performing a hierarchy of timesteps in which individual cells are evolved according to their required timestep criterion as opposed to being evolved alongside all other cells in a global timestep.

We tested the code on an R-type expansion of a ionized bubble in a medium of homogeneous density and found good agreement with the analytical prediction.
We also performed a test in which we studied the formation of a shadow behind an overdense clump.
We found that the method performs well, but that there are significant differences in its behavior compared to the original {\sc Arepo} implementation {\sc Sweep}.
This is due to the different chemistry implementations as well as the slightly different treatment of photon fluxes between the two methods.

We also performed a series of tests investigating the photon-conservation properties of the solver, finding that the fraction of lost photons can be kept comparatively small if appropriate timestep parameters are chosen.

We investigated the performance of the sub-timestepping in a 1D-test in which we studied the equivalent of an R-type expansion.
Since this test is comparatively cheap to run, it allowed us to vary the timestep parameters over a large range of values.
We find that substepping does allow for larger maximum timesteps without sacrificing the accuracy of the solution, which in turns results in a reduction of the overall time to solution.

Our test of the R-type expansion around an ionizing source close to the boundary of the simulation box shows that sub-timesteps help significantly with reducing the overall cost in simulation time that a proper source iteration to convergence incurs. In fact, we find that for our application, the combination of substepping with warmstarting (re-using the fluxes from a previous iteration) is a good approach to reach the desired levels of accuracy in simulations with periodic boundary conditions.

We also briefly discussed and tested the implementation of radiation chemistry in {\sc Subsweep}, which is a simple explicit solver with internal substepping that tracks hydrogen and the corresponding ionization and heating processes.

The primary extension to this method that could further improve the performance is to allow cells to change their timestep level in between full sweeps, in order to quickly react to sudden changes in the physical variables of the cell, something we chose not to do because of the additional complexity that comes along with the implementation.
Doing so could help with increasing the accuracy of the integration (in the case of a suddenly decreasing timestep) and improve performance (in the case of a suddenly increasing timestep).

We believe that this new improvement to the sweep method is a significant step in order to incorporate it into a full cosmological simulation with hydrodynamics, gravity, and a proper treatment of star formation.
In {\sc Sweep}, there is no substepping, such that a global sweep timestep was required. This timestep has to be low enough to enable accurate integration of the radiation chemistry, but decreasing it drastically increases the run-time of the simulation.
Substepping allows the sweep algorithm to perform accurate integration without incurring a prohibitively large computational cost and is therefore a very promising extension to the sweep method.
One challenge with a implementation of sweep substepping into hydrodynamics simulations is that most state-of-the-art cosmological codes already contain their own internal sub-timestep hierarchy, usually for both gravity and hydrodynamics.
The inclusion of the sweep substepping method would require properly integrating with those timestepping schemes, especially the hydrodynamical one, since they interact - radiative transfer can lead to increases in temperature which could lead to a reduced hydrodynamical timestep. Conversely, hydrodynamical interaction can also a sudden increase in the required accuracy of the integration of radiative transfer.
However, despite all of this additional complexity, we believe that integration of substepping into a full hydrodynamical code could potentially yield large benefits by bringing a method that efficiently computes very accurate solutions to the radiative transfer equations to cosmological simulations.

\begin{acknowledgements}
We thank the anonymous referee for highly insightful comments that have helped to improve the manuscript.  We acknowledge computing resources and data storage facilities provided by the State of Baden-W\"{u}rttemberg through bwHPC and the German Research Foundation (DFG) through grant INST 35/1134-1 FUGG and INST 35/1503-1 FUGG. We also thank for computing time from the Leibniz Computing Center (LRZ) in project pr74nu. We thank for funding from the Heidelberg Cluster of Excellence EXC 2181 (Project-ID 390900948) `STRUCTURES: A unifying approach to emergent phenomena in the physical world, mathematics, and complex data' supported by the German Excellence Strategy, from the European Research Council in the ERC synergy grant `ECOGAL – Understanding our Galactic ecosystem: From the disk of the Milky Way to the formation sites of stars and planets' (project ID 855130), and from DFG via the Collaborative Research Center (SFB 881, Project-ID 138713538) 'The Milky Way System' (subprojects A1, B1, B2, B8). We thank Dylan Nelson, Annalisa Pillepich, and Matthew C. Smith for useful discussions.
\end{acknowledgements}



\bibliographystyle{aa} 
\bibliography{library}

\begin{thebibliography}{50}
\expandafter\ifx\csname natexlab\endcsname\relax\def\natexlab#1{#1}\fi

\bibitem[{Adams {et~al.}(2020)Adams, Adams, Hawkins, Smith, Rauchwerger, Amato,
  Bailey, Falgout, Kunen, \& Brown}]{adamsProvablyOptimalParallel2019}
Adams, M.~P., Adams, M.~L., Hawkins, W.~D., {et~al.} 2020, JCP, 407, 109234

\bibitem[{Aubert \& Teyssier(2008)}]{aubertRadiativeTransferScheme2008}
Aubert, D. \& Teyssier, R. 2008, MNRAS, 387, 295

\bibitem[{Baczynski {et~al.}(2015)Baczynski, Glover, \&
  Klessen}]{baczynskiFerventChemistrycoupledIonizing2015}
Baczynski, C., Glover, S. C.~O., \& Klessen, R.~S. 2015, MNRAS, 454, 380

\bibitem[{Baker \& Koch(1998)}]{bakerAlgorithmMassivelyParallel1998}
Baker, R.~S. \& Koch, K.~R. 1998, NSE, 128, 312

\bibitem[{Boss(2008)}]{bossFluxlimitedDiffusionApproximation2008}
Boss, A.~P. 2008, ApJ, 677, 607

\bibitem[{Bowyer(1981)}]{bowyerComputingDirichletTessellations1981}
Bowyer, A. 1981, Comput. J., 24, 162

\bibitem[{Cen(1992)}]{cenHydrodynamicApproachCosmology1992}
Cen, R. 1992, ApJS, 78, 341

\bibitem[{D'Aloisio {et~al.}(2020)D'Aloisio, McQuinn, Trac, Cain, \&
  Mesinger}]{daloisioHydrodynamicResponseIntergalactic2020}
D'Aloisio, A., McQuinn, M., Trac, H., Cain, C., \& Mesinger, A. 2020, ApJ, 898,
  149

\bibitem[{Dullemond {et~al.}(2012)Dullemond, Juhasz, Pohl, Sereshti, Shetty,
  Peters, Commercon, \& Flock}]{dullemondRADMC3DMultipurposeRadiative2012}
Dullemond, C.~P., Juhasz, A., Pohl, A., {et~al.} 2012, ASCL, ascl:1202.015

\bibitem[{Edelsbrunner \&
  Shah(1996)}]{edelsbrunnerIncrementalTopologicalFlipping1996}
Edelsbrunner, H. \& Shah, N.~R. 1996, Algorithmica, 15, 223

\bibitem[{Eldridge {et~al.}(2017)Eldridge, Stanway, Xiao, McClelland, Taylor,
  Ng, Greis, \& Bray}]{eldridgeBinaryPopulationSpectral2017}
Eldridge, J.~J., Stanway, E.~R., Xiao, L., {et~al.} 2017, PASA, 34, e058

\bibitem[{{Finlator} {et~al.}(2018){Finlator}, {Keating}, {Oppenheimer},
  {Dav{\'e}}, \& {Zackrisson}}]{Finlator2018}
{Finlator}, K., {Keating}, L., {Oppenheimer}, B.~D., {Dav{\'e}}, R., \&
  {Zackrisson}, E. 2018, \mnras, 480, 2628

\bibitem[{Gnedin \&
  Abel(2001)}]{gnedinMultidimensionalCosmologicalRadiative2001}
Gnedin, N.~Y. \& Abel, T. 2001, New Astron., 6, 437

\bibitem[{{Gnedin} \& {Madau}(2022)}]{Gnedin2022}
{Gnedin}, N.~Y. \& {Madau}, P. 2022, Living rev. comput. astrophys., 8, 3

\bibitem[{Haiman {et~al.}(1996)Haiman, Thoul, \&
  Loeb}]{haimanCosmologicalFormationLowMass1996}
Haiman, Z., Thoul, A.~A., \& Loeb, A. 1996, ApJ, 464, 523

\bibitem[{Hayes \& Norman(2003)}]{hayesFluxlimitedDiffusionParallel2003}
Hayes, J.~C. \& Norman, M.~L. 2003, ApJS, 147, 197

\bibitem[{Hui \& Gnedin(1997)}]{huiEquationStatePhotoionized1997}
Hui, L. \& Gnedin, N.~Y. 1997, MNRAS, 292, 27

\bibitem[{Iliev {et~al.}(2006)Iliev, Ciardi, Alvarez, Maselli, Ferrara, Gnedin,
  Mellema, Nakamoto, Norman, Razoumov, Rijkhorst, Ritzerveld, Shapiro, Susa,
  Umemura, \& Whalen}]{ilievCosmologicalRadiativeTransfer2006}
Iliev, I.~T., Ciardi, B., Alvarez, M.~A., {et~al.} 2006, MNRAS, 371, 1057

\bibitem[{Jaura {et~al.}(2018)Jaura, Glover, Klessen, \&
  Paardekooper}]{jauraSPRAICouplingRadiative2018}
Jaura, O., Glover, S. C.~O., Klessen, R.~S., \& Paardekooper, J.-P. 2018,
  MNRAS, 475, 2822

\bibitem[{Kannan {et~al.}(2019)Kannan, Vogelsberger, Marinacci, McKinnon,
  Pakmor, \& Springel}]{kannanArepoRTRadiationHydrodynamics2019}
Kannan, R., Vogelsberger, M., Marinacci, F., {et~al.} 2019, MNRAS, 485, 117

\bibitem[{Koch {et~al.}(1991)Koch, Baker, \&
  Alcouffe}]{kochSolutionFirstorderForm1991}
Koch, K.~R., Baker, R.~S., \& Alcouffe, R.~E. 1991, Trans. Am. Nucl. Soc., 65,
  198

\bibitem[{Krumholz {et~al.}(2007)Krumholz, Klein, \&
  McKee}]{krumholzRadiationHydrodynamicSimulationsCollapse2007}
Krumholz, M.~R., Klein, R.~I., \& McKee, C.~F. 2007, ApJ, 656, 959

\bibitem[{{Krumholz} {et~al.}(2007){Krumholz}, {Stone}, \&
  {Gardiner}}]{krumholz2007}
{Krumholz}, M.~R., {Stone}, J.~M., \& {Gardiner}, T.~A. 2007, \apj, 671, 518

\bibitem[{{Leong} {et~al.}(2023){Leong}, {Meiksin}, {Lai}, \& {To}}]{kahou2023}
{Leong}, K.-H., {Meiksin}, A., {Lai}, A., \& {To}, K.~H. 2023, \mnras, 519,
  5743

\bibitem[{Levermore \&
  Pomraning(1981)}]{levermoreFluxlimitedDiffusionTheory1981}
Levermore, C.~D. \& Pomraning, G.~C. 1981, ApJ, 248, 321

\bibitem[{Loeb \& Barkana(2001)}]{loebReionizationUniverseFirst2001}
Loeb, A. \& Barkana, R. 2001, Annu. Rev. Astron. Astrophys., 39, 19

\bibitem[{{Marinacci} {et~al.}(2018){Marinacci}, {Vogelsberger}, {Pakmor},
  {Torrey}, {Springel}, {Hernquist}, {Nelson}, {Weinberger}, {Pillepich},
  {Naiman}, \& {Genel}}]{marinacciFirstResultsIllustrisTNG2018}
{Marinacci}, F., {Vogelsberger}, M., {Pakmor}, R., {et~al.} 2018, \mnras, 480,
  5113

\bibitem[{Mihalas \&
  {Weibel-Mihalas}(1999)}]{mihalasFoundationsRadiationHydrodynamics1999}
Mihalas, D. \& {Weibel-Mihalas}, B. 1999, Foundations of Radiation
  Hydrodynamics ({Mineola, N.Y}: {Dover})

\bibitem[{Naiman {et~al.}(2018)Naiman, Pillepich, Springel, {Ramirez-Ruiz},
  Torrey, Vogelsberger, Pakmor, Nelson, Marinacci, Hernquist, Weinberger, \&
  Genel}]{naimanFirstResultsIllustrisTNG2018}
Naiman, J.~P., Pillepich, A., Springel, V., {et~al.} 2018, MNRAS, 477, 1206

\bibitem[{Nelson {et~al.}(2015)Nelson, Pillepich, Genel, Vogelsberger,
  Springel, Torrey, Rodriguez-Gomez, Sijacki, Snyder, Griffen, Marinacci,
  Blecha, Sales, Xu, \& Hernquist}]{nelsonIllustrisTNGSimulationsPublic2021}
Nelson, D., Pillepich, A., Genel, S., {et~al.} 2015, Astron. Comput., 13, 12

\bibitem[{Nelson {et~al.}(2019)Nelson, Pillepich, Springel, Pakmor, Weinberger,
  Genel, Torrey, Vogelsberger, Marinacci, \&
  Hernquist}]{nelsonFirstResultsTNG502019}
Nelson, D., Pillepich, A., Springel, V., {et~al.} 2019, MNRAS, 490, 3234

\bibitem[{Nelson {et~al.}(2018)Nelson, Pillepich, Springel, Weinberger,
  Hernquist, Pakmor, Genel, Torrey, Vogelsberger, Kauffmann, Marinacci, \&
  Naiman}]{nelsonFirstResultsIllustrisTNG2018}
Nelson, D., Pillepich, A., Springel, V., {et~al.} 2018, MNRAS, 475, 624

\bibitem[{Osterbrock \&
  Ferland(2006)}]{osterbrockAstrophysicsGaseousNebulae2006}
Osterbrock, D.~E. \& Ferland, G.~J. 2006, Astrophysics of Gaseous Nebulae and
  Active Galactic Nuclei (University Science Books)

\bibitem[{Oxley \& Woolfson(2003)}]{oxleySmoothedParticleHydrodynamics2003}
Oxley, S. \& Woolfson, M.~M. 2003, MNRAS, 343, 900

\bibitem[{Peter {et~al.}(2023)Peter, Klessen, Kanschat, Glover, \&
  Bastian}]{peterSweepMethodRadiative2023}
Peter, T., Klessen, R.~S., Kanschat, G., Glover, S. C.~O., \& Bastian, P. 2023,
  MNRAS, 519, 4263

\bibitem[{Pillepich {et~al.}(2018)Pillepich, Nelson, Hernquist, Springel,
  Pakmor, Torrey, Weinberger, Genel, Naiman, Marinacci, \&
  Vogelsberger}]{pillepichFirstResultsIllustrisTNG2018}
Pillepich, A., Nelson, D., Hernquist, L., {et~al.} 2018, MNRAS, 475, 648

\bibitem[{Pillepich {et~al.}(2019)Pillepich, Nelson, Springel, Pakmor, Torrey,
  Weinberger, Vogelsberger, Marinacci, Genel, {van der Wel}, \&
  Hernquist}]{pillepichFirstResultsTNG502019}
Pillepich, A., Nelson, D., Springel, V., {et~al.} 2019, MNRAS, 490, 3196

\bibitem[{{Robitaille}(2011)}]{hyperion}
{Robitaille}, T.~P. 2011, \aap, 536, A79

\bibitem[{Rosdahl {et~al.}(2013)Rosdahl, Blaizot, Aubert, Stranex, \&
  Teyssier}]{rosdahlRamsesrtRadiationHydrodynamics2013}
Rosdahl, J., Blaizot, J., Aubert, D., Stranex, T., \& Teyssier, R. 2013, MNRAS,
  436, 2188

\bibitem[{Rybicki \&
  Lightman(1985)}]{rybickiRadiativeProcessesAstrophysics1985}
Rybicki, G.~B. \& Lightman, A.~P. 1985, Radiative {{Processes}} in
  {{Astrophysics}} ({Weinheim, Germany}: {Wiley-VCH Verlag GmbH \& Co. KGaA})

\bibitem[{Shapiro {et~al.}(2004)Shapiro, Iliev, \&
  Raga}]{shapiroPhotoevaporationCosmologicalMinihaloes2004}
Shapiro, P.~R., Iliev, I.~T., \& Raga, A.~C. 2004, MNRAS, 348, 753

\bibitem[{Springel(2010)}]{springelPurSiMuove2010}
Springel, V. 2010, MNRAS, 401, 791

\bibitem[{Springel {et~al.}(2018)Springel, Pakmor, Pillepich, Weinberger,
  Nelson, Hernquist, Vogelsberger, Genel, Torrey, Marinacci, \&
  Naiman}]{springelFirstResultsIllustrisTNG2018}
Springel, V., Pakmor, R., Pillepich, A., {et~al.} 2018, MNRAS, 475, 676

\bibitem[{Str{\"o}mgren(1939)}]{stromgrenPhysicalStateInterstellar1939}
Str{\"o}mgren, B. 1939, ApJ, 89, 526

\bibitem[{Vermaak {et~al.}(2020)Vermaak, Ragusa, Adams, \&
  Morel}]{vermaakMassivelyParallelTransport2020}
Vermaak, J. I.~C., Ragusa, J.~C., Adams, M.~L., \& Morel, J.~E. 2020, J.
  Comput. Phys., 425, 109892

\bibitem[{Watson(1981)}]{watsonComputingNdimensionalDelaunay1981}
Watson, D.~F. 1981, Comput. J., 24, 167

\bibitem[{Whitehouse \&
  Bate(2004)}]{whitehouseSmoothedParticleHydrodynamics2004}
Whitehouse, S.~C. \& Bate, M.~R. 2004, MNRAS, 353, 1078

\bibitem[{Wise(2019)}]{wiseCosmicReionisation2019}
Wise, J.~H. 2019, Contemp. Phys., 60, 145

\bibitem[{Zaroubi(2013)}]{zaroubiEpochReionization2013}
Zaroubi, S. 2013, in The {{First Galaxies}}, ed. T.~Wiklind, B.~Mobasher, \&
  V.~Bromm, Vol. 396 ({Berlin, Heidelberg}: {Springer Berlin Heidelberg}),
  45--101

\bibitem[{Zeyao \& Lianxiang(2004)}]{zeyaoParallelFluxSweep2004}
Zeyao, M. \& Lianxiang, F. 2004, J. Supercomput., 30, 5

\end{thebibliography}

\appendix
\section{Details of the radiation chemistry}
\label{sec:subsweep-appendix-radchem}
Here, we specify the exact equations used in our radiation chemistry solver described in \sref{chemistry}.
The photo-heating term $H_{\mathrm{photo}}$ is given by
\begin{align}
  H_{\mathrm{photo}} &= R \bb{1 - e^{-\neutralHydrogenDensity \sigma l}} (E_{\rm avg} - E_{\rm Ryd}),
\end{align}
where $R$ is the rate per unit volume at which ionizing photons enter the cell, $l$ is the size of the cell, $E_{\rm Ryd} = \SI{13.65}{\eV}$ is the Rydberg energy and $E_{\rm avg}$ is the number-averaged photon energy defined as 
\begin{align}
    E_{\rm avg} = \frac{\int_{f_{\mathrm{ion}}}^{\infty} J_{\nu} \; \mathrm{d} \nu }{\int_{f_{\mathrm{ion}}}^{\infty} J_{\nu} / h \nu \; \mathrm{d} \nu },
\end{align}
which, under the assumption of the spectrum of the BPASS \citep{eldridgeBinaryPopulationSpectral2017} source model, becomes $E_{\rm avg}= \SI{18.028}{\eV}$.
Collisional ionization rates $\beta(T)$, collisional ionization cooling rates $\zeta(T)$ and
collisional excitation rates $\psi(T)$ are given by \citep{cenHydrodynamicApproachCosmology1992}:
\begin{align}
  \beta(T) &= \SI{5.85e-11}{\cubic \cm \per \second} \sqrt{T / (1K)} \\
  \nonumber &\bb{1 + \sqrt{T / \bb{\SI{e5}{\kelvin}}}}^{-1} e^{-\SI{157809.1}{\kelvin} / T}, \\
  \zeta(T) &= \SI{1.27e-21}{\erg \cubic \cm \per \second} \sqrt{T / (1K)} \\
  \nonumber &\bb{1 + \sqrt{T / \bb{\SI{e5}{\kelvin}}}}^{-1} e^{-\SI{157809.1}{\kelvin} / T}, \\
  \psi(T) &= \SI{7.5e-19}{\erg \cubic \cm \per \second} \sqrt{T / (1K)} \\
  \nonumber &\bb{1 + \sqrt{T / \bb{\SI{e5}{\kelvin}}}}^{-1} e^{-\SI{118348}{\kelvin} / T}.
\end{align}
As discussed in \sref{chemistry}, we use the on-the-spot approximation and therefore only consider case B recombination. The case B recombination rates $\alpha(T)$ and recombination cooling rates $\eta(T)$ are given by \citep{huiEquationStatePhotoionized1997}
\begin{align}
  \alpha(T) &= \SI{2.753e-14}{\cubic \cm \per \second} T \frac{\lambda^{1.5}}{\bb{1 + (\lambda / 2.74)^{0.407}}^{2.242}}, \\
  \eta(T) &= \SI{3.435e-30}{\erg \cubic \cm \per \second \per \kelvin} T \frac{\lambda^{1.97}}{\bb{1 + (\lambda / 2.25)^{0.376}}^{3.72}},
\end{align}
where $\lambda = \SI{315614}{\kelvin} / T$.
The bremsstrahlung cooling rate coefficient $\Theta(T)$ is given by \citep{osterbrockAstrophysicsGaseousNebulae2006}:
\begin{align}
  \Theta(T) &= \SI{1.42e-27}{\erg \cubic \cm \per \second} \sqrt{T / (\SI{1}{\kelvin})}.
\end{align}
Compton cooling $\bar{\omega}(T)$ is defined as \citep{haimanCosmologicalFormationLowMass1996}
\begin{align}
  \bar{\omega}(T) &= \SI{1.017e-37}{\erg \per \second} \bb{\frac{2.727}{a}}^4 \bb{T-\frac{2.727}{a}},
\end{align}
where $\SI{2.727}{\kelvin} / a$ is the CMB temperature and $a$ is the cosmological scale factor.
Our chemistry solver also uses the temperature derivatives of the rate coefficients, which are obtained by symbolic differentiation.

The value of the heating and cooling rates as a function of temperature, for a fixed density of $\rho = m_{\mathrm{p}} \cdot \SI{0.01}{\per \cubic \centi \meter}$, where $m_{\mathrm{p}}$ is the proton mass and a fixed hydrogen ionization fraction of $x_{\mathrm{HII}} = 0.5$ are shown in \fref{chemistry_rates}.
\begin{figure}
    \centering
    \includegraphics[width=\columnwidth]{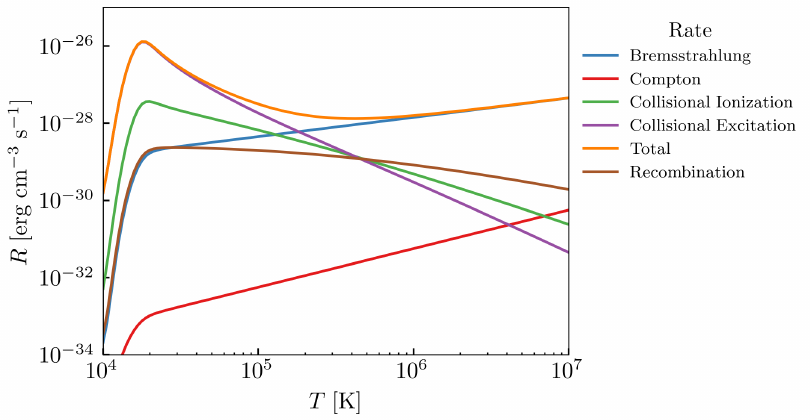}
    \caption{\label{fig:chemistry_rates}Value of the different cooling rates as a function of temperature for $\rho = m_{\mathrm{p}} \cdot \SI{0.01}{\per \cubic \centi \meter}$, and $x_{\mathrm{HII}} = 0.5$.
    }
\end{figure}

\end{document}